\documentclass{aa}  

\usepackage{graphicx}
\usepackage{txfonts}
\usepackage{verbatim}
\usepackage{subcaption}
\usepackage[normalem]{ulem}
\usepackage{xcolor}

\newcommand*{\msun}{\ensuremath{\rm{M}_{\odot}}\text{ }}

\newcommand*    \kpc{{\,\mathrm{kpc}}}

\newcommand*    \gyr{{\,\rm Gyr}}

\begin{document}

   \title{Realistic consecutive galaxy mergers form eccentric PTA sources}

   %\subtitle{Time-scales and black hole triplets formation}

   \author{F. Fastidio
          \inst{1,2,4}\fnmsep\thanks{f.fastidio@campus.unimib.it}
          \and
          E. Bortolas\inst{5}
          \and
          A. Gualandris\inst{2}
          \and
          A. Sesana\inst{1,3,4}
          \and
          J. I. Read\inst{2}
          \and
          W. Dehnen\inst{6}
          }

   \institute{Dipartimento di Fisica ``G. Occhialini", Universit{\'a} degli Studi di Milano-Bicocca, Piazza della Scienza 3, I-20126 Milano,            Italy\label{unimib}\\
             %\email{wuchterl@amok.ast.univie.ac.at}
             \and
             School of Mathematics and Physics, University 
             of Surrey, Guildford GU2 7XH, UK\label{surrey}\\
             %\email{c.ptolemy@hipparch.uheaven.space}
             %\thanks{The university of heaven temporarily does not accept e-mails}
             \and
             INAF - Osservatorio Astronomico di Brera, via Brera 20, I-20121 Milano, Italy\label{inaf-brera}\\
             \and
             INFN, Sezione di Milano-Bicocca, Piazza della Scienza 3, I-20126 Milano, Italy\label{infn-unimib}\\
             \and
             INAF - Osservatorio Astronomico di Padova, Vicolo dell’Osservatorio, 5, I-35122 Padova (PD), Italy\label{inaf-padova}\\
             \and
             Astronomisches Rechen-Institut, Zentrum f{\"u}r Astronomie der Universit{\"a}t Heidelberg, M{\"o}nchhofstra\ss{}e 12-14, 69120, Heidelberg, Germany
             }

   \date{Received Month DD, YYYY; accepted Month DD, YYYY}

%\abstract{}{}{}{}{} 
% 5 {} token are mandatory
 
\abstract
{Results from pulsar timing arrays (PTAs) show evidence of a gravitational wave background (GWB) consistent with a population of unresolved supermassive black hole binaries (BHBs). The observed spectrum shows a flattening at lower frequencies that can be explained by a population of eccentric BHBs. This study aims to determine the dynamical evolution and merger timescales of the most massive BHBs, which are potential sources of the GWB. We select successive galactic major mergers from the IllustrisTNG100-1 cosmological simulation and re-simulate them at high resolution with the $N$-body code {\tt Griffin}, down to binary separations of the order of a parsec. Coalescence timescales are estimated using a semi-analytical model that incorporates gravitational wave emission and stellar hardening. Throughout our investigation, we consider the impact of prior mergers on the remnant galaxy in the form of core scouring and anisotropy, which can influence the subsequent formation and evolution of BHBs. We find that all the binaries in our sample enter the PTA band with an eccentricity $e>0.85$: such a large eccentricity can impact the shape of the PTA observed GWB spectrum, and it highlights the importance of including the eccentricity of binaries when interpreting the PTA signal. Furthermore, we find that: (i) starting from initial separations of a few tens of kpcs, the dynamical friction phase lasts for a few hundred Myrs; (ii) the binary formation time is not resolution dependent; (iii) the scatter on the eccentricity at binary formation decreases with increasing resolution; (iv) triple systems form whenever a third galaxy interacts with a binary which hasn't yet reached coalescence.
}


   \maketitle
%
%-------------------------------------------------------------------

\section{Introduction}
   %(\citeyear{baker}) standard one-zone model. 
Supermassive black holes (BHs) are expected to reside at the centre of massive galaxies, with masses ranging between $10^{6}-10^{10}$ \msun \citep[e.g.][]{1995ARA&A..33..581K, 2013ARA&A..51..511K}. The $\Lambda$CDM theoretical model for galaxy evolution predicts that structures in the Universe form hierarchically. In this process, smaller systems develop first, followed by the formation of larger structures through accretion and subsequent mergers \citep{1977ApJ...217L.125O, 1980MNRAS.191P...1W, 1993MNRAS.262..627L}. A natural result of these galactic encounters is the formation of BH pairs, where each black hole resides in the merger remnant at a separation of order $\kpc{}$ from its companion \citep{1992ARA&A..30..705B}. The dynamical evolution of the two BHs is governed by various physical processes, that extract energy and angular momentum, drawing them toward the centre of the remnant, and potentially leading to their coalescence \citep{1980Natur.287..307B}. 

The first mechanism that contributes to the slowing down of the BHs is dynamical friction against the stellar and dark matter (DM) content of the host galaxy \citep{1943ApJ....97..255C}. Dynamical friction ceases to be efficient when the velocity of the black holes becomes comparable with the velocity of the surrounding stars, and this occurs around the time the two BHs form a bound binary system, which typically happens at separations of about 100 pc for a $10^9$ \msun binary.

From that point onward, assuming a gas-poor environment, the binary shrinks (`hardens') due to 3-body interactions with single stars: they eventually get ejected at high velocities, carrying away kinetic energy and thus increasing the binding energy of the black hole binary \citep[BHB,][]{1996NewA....1...35Q, 2006ApJ...651..392S}.  This mechanism is expected to be efficient due to the triaxiality of the host's potential that is induced by the merger process \citep{2016ApJ...828...73K, 2018MNRAS.477.2310B}. In triaxial potentials, the total angular momentum of stars is not conserved along the orbit, allowing for a large population of stars on centrophilic orbits. These may pass arbitrarily close to the centre, where they can interact with the central BHB \citep[e.g.][]{2017MNRAS.464.2301G}.

Assuming that stellar hardening is capable of driving a $\sim 10^9$ \msun BHB to a separation of the order of 1--0.1 pc on time-scales shorter than a Hubble time, the physical process responsible for the final stages of binary evolution is gravitational wave (GW) emission, which can efficiently bring the system to coalescence \citep{1964PhRv..136.1224P}.

The superposition of the GW signals produced by a population of unresolved BHBs is expected to result in a gravitational wave background \citep[GWB;][]{1995ApJ...446..543R, 2003ApJ...583..616J,2008MNRAS.390..192S}. The GWB can be observed via pulsar timing arrays \citep[PTAs;][]{1990ApJ...361..300F} by searching for correlated deviations in the times of arrival (TOAs) of radio signals coming from millisecond pulsars. PTAs are mostly sensitive to low-redshift ($z<2$), massive sources ($M>10^8$ \msun, with mass ratio $q>1/4$), which are expected to dominate the GWB signal \citep[e.g.][]{2008MNRAS.390..192S,2022MNRAS.509.3488I}. 
In summer 2023, the European PTA (EPTA) and Indian PTA (InPTA), The North America Nanohertz Observatory for GWs (NANOGrav), the Parkes PTA (PPTA) and the Chinese PTA (CPTA) reported evidence for a correlated GWB, with statistical significance between 2-4 $\sigma$ \citep{2023A&A...678A..48E,2023A&A...678A..49E,2023A&A...678A..50E,2024A&A...685A..94E,2023arXiv230616226A,2023PhRvL.131q1001S,2023ApJ...951L...9A,2023ApJ...951L..10A,2023ApJ...951L..50A,2023ApJ...951L...8A,2023ApJ...951L..11A,2023ApJ...951L...6R,2023RAA....23g5024X}.

The amplitude of the GWB depends on the number of sources emitting GWs within the frequency range that PTAs are sensitive to ($10^{-9}-10^{-7}$ Hz). To be effective GW emitters, BHBs must achieve separations on the order of 1--0.1 pc (see above). For this reason, studying the timescales required for BHBs to reach this separation is essential for interpreting the GWB signal. Moreover, depending on the galactic environment and the efficiency of each physical process that drives the dynamical evolution of the binary, the lifetime of the BHB can exceed the time elapsing between a galactic merger and the subsequent one. As a result, systems including more than two BHs may form \citep[e.g.][]{2007MNRAS.377..957H, 2025arXiv250604369S}. According to  \cite{2006ApJ...640..241B} and \cite{2008ApJ...681..232L}, $\sim30-70\%$ of present-day spheroidal galaxies have experienced a major merger since $z=1$. Thus, on average, all massive galaxies have undergone a merger in the last $10\gyr{}$, leading to $\sim 10\%$ of BHBs to possibly form a triplet \citep[assuming their lifetime to be $\sim1\gyr{}$,][]  {2010MNRAS.402.2308A}.  

This paper has the main goal of simulating, for the first time,  realistic consecutive major mergers of massive galaxies at low redshift, which can produce PTA sources. Starting from cosmological initial conditions, we accurately follow the dynamics of the BHBs thus formed using the state-of-the-art Fast Multiple Method (FMM) code {\tt Griffin} \citep{2014ComAC...1....1D}, with the goal of determining the evolutionary time scales and the properties of these system, including the binary eccentricity and the the probability of forming triplets.

The paper is structured as follows. In Section \ref{sec:methods} we describe the methods, reporting (i) the selection criteria for merger trees in our
chosen cosmological simulation (\ref{sec:TNG}); (ii) the method adopted to initialise the first merger of the tree (\ref{IC primo merger}); (iii) the precautions taken when initialising subsequent mergers (\ref{IC following mergers}) and (iv) the methods used to evolve the systems down to coalescence \ref{sec:time evol}. We present our results in Section \ref{Results}, discussing and comparing them with previous works. Finally, in Section \ref{Conclusions} we draw our conclusions.

%--------------------------------------------------------------------
\section{Methods} \label{sec:methods}

\subsection{Selecting merger trees in IllustrisTNG100-1} \label{sec:TNG}

 IllustrisTNG100-1 (hereafter TNG100-1) is the highest resolution cosmological, magneto-hydrodynamical simulation in the IllustrisTNG100 simulation suite \citep{2018MNRAS.475..676S, 2018MNRAS.475..648P, 2018MNRAS.475..624N, 2018MNRAS.477.1206N, 2018MNRAS.480.5113M}. TNG100-1 simulates a cubic volume with side length of $110.7\,{\rm Mpc}$, using AREPO \citep{2010MNRAS.401..791S}, a moving, unstructured-mesh hydrodynamic code, with superposed smoothed-particle hydrodynamics (SPH) gas particles \citep[e.g.][]{2010ARA&A..48..391S}. Star and DM particles have masses, respectively,  $m_{\rm{baryon}}=1.4\times 10^6 \msun$ and $m_{\rm{DM}}=7.5\times 10^6 \msun$ and softening length $\varepsilon=0.74$ kpc. BHs are seeded with a mass $M_{\rm{seed}}=8 \times 10^5 h^{-1}\msun$ in halos with $M_{\rm{h}}\geq5 \times 10^{10}h^{-1}\msun$. Their accretion follows the Bondi-Hoyle accretion model \citep{1944MNRAS.104..273B} capped at the Eddington limit; they are dynamically evolved, while their position is kept fixed at the potential minimum of the host galaxy. 
 
All TNG100-1 data are stored in 99 snapshots, publicly available at www.tng-project.org. The merger history of the structures is easily reconstructable through the provided merger trees, built using the {\tt SubLink} algorithm \citep{2015MNRAS.449...49R}. In our work, we consider two galaxies to merge when they share the same {\tt SubLink} Descendant, and we define the merger snapshot as the snapshot of said Descendant.

\begin{table*}
    \centering
    \begin{tabular}{ccccccc}
         \# merger&  z&  M$_{\rm{BH}}$ [M$_{\odot}$]&  M$_{\rm{b}}$ [M$_{\odot}$]&  M$_{\rm{h}}$ [M$_{\odot}$]&  e& peri [kpc]\\
         \hline\hline
         \multicolumn{7}{c}{ID merger tree = 17187}\\
         \hline\hline
         I&  1.90&  4.23e+08&  9.89e+10&  2.75e+12&  0.99& 2.71e-01\\
 & & 2.19e+08& 4.09e+10& 3.84e+10& &\\
 \hline
         II&  0.21&  7.29e+08&  5.34e+11&  1.29e+14&  0.39& 3.70e+01\\
 & & 7.21e+08& 1.27e+11& 2.87e+11& &\\
 \hline
         III&  0.17&  (7.29+7.21)e+08&  (5.34+1.27)e+11&  (1.29+2.87)e+14&  0.89& 3.64\\
 & & 7.30e+08& 8.94e+10& 1.24e+11& &\\
 \hline\hline
         \multicolumn{7}{c}{ID merger tree = 125027}\\
         \hline\hline
         I&  2.00&  4.64e+08&  9.57e+10&  3.85e+12&  0.985& 9.17e-02\\
 & & 1.26e+08& 2.51e+10& 1.06e+10& &\\
 \hline
         II&  1.60&  5.97e+08&  1.41e+11&  6.75e+12&  0.97& 1.32\\
 & & 3.32e+08& 3.91e+10& 1.59e+11& &\\
 \hline\hline
         \multicolumn{7}{c}{ID merger tree = 125028}\\
         \hline\hline
 I& 1.60& 3.16e+08& 6.63e+10& 3.15e+11& 0.98&1.09\\
 & & 3.55e+08& 7.73e+10& 4.71e+12& &\\
 \hline
 II& 1.41& (3.16+3.55)e+08& (6.63+7.73)e+10& (0.315+4.71)e+12& 0.85&7.63e-01\\
 & & 3.55e+08& 3.22e+10& 7.25e+09& &\\
 \hline
 III& 0.92& 6.91e+08& 3.16e+11& 1.29e+13& 0.987&2.99e-01\\
 & & 6.49e+08& 1.06e+11& 6.93e+10& &\\
 \hline\hline
 \multicolumn{7}{c}{ID merger tree = 168390}\\
 \hline\hline
 I& 1.41& 3.31e+08& 4.74e+10& 4.17e+10& 0.775&9.26\\
 & & 2.22e+08& 7.44e+10& 5.71e+12& &\\
 \hline
 II& 1.30& (3.31+2.22)e+08& (4.74+7.44)e+10& (0.04+5.71)e+12& 0.947&3.23e-01\\
 & & 3.26e+08& 3.53e+10& 7.66e+09& &\\
 \hline\hline
 \multicolumn{7}{c}{ID merger tree = 197109}\\
 \hline\hline
 I& 2.00& 2.49e+07& 1.998e+10& 1.07e+12& 0.745&2.25\\
 & & 2.26e+07& 5.24e+09& 6.75e+09& &\\
 \hline
 II& 1.36& 1.27e+08& 6.61e+10& 2.50e+12& 0.986&2.35e-01\\
 & & 1.25e+08& 2.68e+10& 3.39e+10& &\\
 \hline
 III& 1.30& (1.27+1.25)e+08& (6.61+2.68)e+10& (0.03+2.50)e+12& 0.997&2.22e-01\\
 & & 1.46e+08& 1.78e+10& 3.46e+09& &\\
    \end{tabular}
    \caption{Parameters related to the mergers in the five merger trees: $z$ is the initial redshift; $M_{\rm{BH}}$, $M_{\rm{b}}$, $M_{\rm{h}}$ denote the mass of the BH, the stellar bulge and the DM halo, respectively. Each box contains data corresponding to the primary (top row) and secondary galaxy (bottom row). When a subsequent merger occurs before the previous merger's remnant has time to form, we report the masses of the primary components (BH, bulge and halo) as the sum of the elements of the previous merger.}
    \label{tab:mergers params}
\end{table*}

To choose our sample of subsequent mergers, we select the last TNG100-1 snapshot ($z=0$) and search for galaxies with stellar mass $M_* \geq 3 \times 10^{11} \msun$ (100 in total). Then we extract their merger trees and we retain only the ones containing more than one major merger ($q\gtrsim 1/4$) at $z<2$. This choice is motivated by the fact that the GWB signal is dominated by sources at low redshift \citep[e.g.][]{2008MNRAS.390..192S,2022MNRAS.509.3488I} and high mass ratio. We obtain a total of 5 merger trees, containing 2 or 3 mergers each. The properties of the selected trees, i.e. (i) the redshift of the mergers, (ii) the BHs masses, (iii) the masses of the stellar and DM components of the progenitor galaxies, and (iv) the orbital parameters, are presented in Table \ref{tab:mergers params}.

\subsection{Creating initial conditions for the first merger in the tree}\label{IC primo merger}

We draw data relative to the progenitor galaxies of each encounter, choosing the TNG100-1 snapshot immediately before the merger\footnote{Due to how the {\tt SubLink} algorithm works, there are instances in which we cannot find both progenitor galaxies in the snapshot before the merger: in this case we follow the two back in time, until both are identified in the same snapshot.} , and we then generate the initial conditions necessary to re-simulate the dynamical evolution of these mergers at high accuracy.

{\tt Griffin} is an FMM \textit{N}-body code, that allows to control the (maximum) force errors directly and reliably, i.e. does not generate an error distribution with a tail towards large force errors, as is otherwise typical for tree/FMM codes \citep{2014ComAC...1....1D}.

We determine the orbit of the merging galaxies by treating them as a Keplerian two-body system.  Since the DM halos may already be overlapping in the selected TNG100-1 snapshot, we consider an effective mass for the computation of the relative orbit, as detailed in \cite{2024MNRAS.532..295F}.

We re-model the merging galaxies, using the action-based galaxy modelling software {\tt AGAMA} \citep{2019MNRAS.482.1525V} to generate the initial conditions. Both the stellar bulge and the DM halo are generated by fitting a Hernquist profile \citep{1990ApJ...356..359H}  to the density profiles in the TNG100-1 snapshot, including an exponential cut-off to match the profile truncation radius.

We initialise the primary galaxy with two different resolutions: \( N = 10^6 \) (low resolution, LR from hereafter) and \( N =2\times 10^6 \) particles (high resolution, HR from hereafter), allocating half of them to the dark matter halo and the remaining half to the stellar bulge. To improve the resolution in the central region of the galaxy, we employ the mass refinement scheme described in \cite{2024MNRAS.529.2150A}. The scheme divides stellar and DM particles into four radial shells each, and oversamples particles in the inner zones at the expense of those in the external regions. The total number of particles is initially increased by up to a factor $10$, then the added particles are retained in the central shell, while the scheme progressively removes particles moving outward, proportionally increasing the mass of the remaining ones. The total mass and density profile are thus preserved.

We determine the number of particles in the secondary galaxy so that: (i) their mass matches that of the particles in the primary galaxy, and (ii) the bulge-halo and primary-secondary mass ratios are maintained. The same mass refinement scheme is then applied to the secondary galaxy as well. To assess the resolution in the innermost part of our models, we calculate the number of particles within 5 times the influence radius of the primary BH\footnote{This is computed on the initial conditions of the first merger of each merger tree, and used as a reference to compare our work with previous literature.} (see Table \ref{tab:res}), defined as the radius enclosing a mass in stars equal to twice the BH mass.

\begin{table*}
    \centering
    \begin{tabular}{c|c|c|c|c}
         ID merger tree&  N($<5$r$_{\rm{inf}}$) - LR &m$_{\rm{BH}}$/m$_{*}$ - LR& N($<5$r$_{\rm{inf}}$) - HR &m$_{\rm{BH}}$/m$_{*}$ - HR\\
         \hline
         17187&  261467&21392.6& 523021&42785.2\\
         125027&  279200&24236.9& 558279&48473.8\\
         125028&  270623&22935.4& 540984&45870.9\\
         168390&  205663&14892.5& 411478&29785.1\\
         197109&  117610&6240.8& 234907&12481.7\\
    \end{tabular}
    \caption{Resolution of our simulations. For each merger tree, we report: (i) the number of stellar and DM particles within 5 influence radii (N($<5$r$_{\rm{inf}}$)), computed for the primary galaxy of the first merger of each merger tree; (ii) the ratio between the primary BH mass and the mass of star particles in the innermost region (m$_{\rm{BH}}$/m$_{*}$).}
    \label{tab:res}
\end{table*}

Improving the resolution comes at the cost of increasing the mass ratio between the DM and stellar particles, which in general is not advisable, since it can lead to the segregation of artificially massive particles towards the centre. To mitigate this effect, we use individual softenings proportional to the particles mass (for both stellar and DM particles):
\begin{equation}
    \varepsilon_{\rm{std}} = \alpha m_{\rm{std}}^{1/3}
\end{equation}
where $m_{\rm std}$ is the particle mass and $\alpha$ is given by
\begin{equation}
    \alpha = \frac{\varepsilon_{0,\rm{std}}}{m_{\rm{h,sh1}}^{1/3}}
\end{equation}
with $\varepsilon_{0,\rm{std}}= 30$ pc and $m_{\rm{h,sh1}}$ is the mass of halo particles in the innermost shell. Similarly, the softening length for BH-BH, BH-stars and BH-dark matter interactions is: 
\begin{equation}
    \varepsilon_{\rm{BH}}=\frac{\varepsilon_{0,\rm{BH}}}{m_{\rm{BH_1}}^{1/3}}m_{\rm{BH}}^{1/3}
    \label{soft_bh}
\end{equation}
where $\varepsilon_{0,\rm{BH}}=1$ pc, $m_{\rm{BH_1}}$ is the mass of the primary BH and $m_{\rm{BH}}$ is the mass of the BH under consideration.

After establishing stable models for both progenitors separately, we position them on the previously computed orbit, with an initial separation equal to the distance between the galactic centres in TNG100-1.

\subsection{Creating initial conditions for the following mergers}\label{IC following mergers}
When initialising the second (or third) merger within the same merger tree, it is important to take some additional precautions. There are two main scenarios to consider: (i) A sufficient amount of time has elapsed between the two mergers, resulting in a well-defined merger remnant and allowing the first BHB to merge; (ii) the time between consecutive mergers is so short that the third galaxy begins to interact with the first two before their merger is completed.

In the first scenario, we manually merge the two BHs when the binary separation reaches the softening length $\varepsilon_{0,\rm{BH}}$. We then allow the system to evolve in isolation until the time at which the second merger is flagged in TNG100-1. However, an inconsistency may arise between the mass of the remnant of the merger in {\tt Griffin} and that of TNG100-1. In {\tt Griffin}, the mass of the remnant is calculated as the sum of the masses of the two progenitors and remains constant over time. In contrast, masses in TNG100-1 evolve over time, which is typical of cosmological hydrodynamic simulations that allow for gas accretion onto the central BHs. Moreover, since galaxies do not evolve in isolation, they can interact with their environment, accrete material, and convert gas into stars. As a result, when initialising the second and third merger, the mass of the primary galaxy (i.e. the merger remnant from the previous merger) in {\tt Griffin} is usually lower than that in TNG100-1. We therefore adjust the galaxy masses to take growth into account, using the same procedure outlined in Sec. \ref{IC primo merger}.

Because scouring in the first major merger carves a core in the stellar distribution, we set up following mergers with a shallower density profile. In particular, we fit the stellar and DM density profiles of the galaxy in {\tt Griffin} with a Dehnen profile \citep{1993MNRAS.265..250D} with central slope $\gamma=0.5$. This has been shown to be consistent with core scouring profiles \citep[e.g.][]{2021MNRAS.502.4794N}.

However, it is important to note that in this work we do not include natal kicks after coalescence, which can lead to the formation of flatter and larger cores, as shown in \cite{2024ApJ...974..204K}. 
When initialising the galaxy with AGAMA, we set the masses of the galaxy components based on values drawn from TNG100-1, but distributed according to the density profiles thus obtained from {\tt Griffin}.

Additionally, we compute the anisotropy parameter $\beta$ from the {\tt Griffin}'s last snapshot as:
\begin{equation}
    \beta = 1-\frac{\overline{v_{\theta}^{2}}+\overline{v_{\phi}^{2}}}{2\overline{v_{\rm{r}}^{2}}}
    \label{beta param}
\end{equation}
where $\overline{v_{\theta}^{2}}$,  $\overline{v_{\phi}^{2}}$ and $\overline{v_{\rm{r}}^{2}}$ are the velocity dispersions in the three components of the spherical coordinates. We then plot the radial anisotropy profile, excluding the regions where the number of particles is $<1000$. The resulting profiles are typically nearly flat, with $-0.29< \beta < -0.24$. We use this constant value of $\beta$ in AGAMA to incorporate the anisotropy into the remnant model that we generate. 

In the second scenario described above, where the third galaxy begins to interact before the previous merger is complete, we consider the inclination angle between the trajectory of the new incoming galaxy and the orbital plane of the progenitors from the previous merger. In the snapshot before the new merger in TNG100-1, the previous merger has already formed a remnant. Note that the resolution of the cosmological simulation is significantly lower than that of the {\tt Griffin} simulation. The third galaxy, on the other hand, is approaching along an orbit that can be defined as outlined in Sec. \ref{IC primo merger}, and whose total angular momentum can be easily calculated. We adopt the total angular momentum of the stellar component of the remnant galaxy in TNG100-1 (i.e. the primary progenitor of the new merger) as a good approximation for the total angular momentum of the orbit of the previous merger. We then compute the relative inclination from the two angular momenta and set up the inclination between the first and the second merger using this value.

\subsection{Time evolution}\label{sec:time evol}
For each merger tree, we initialise the first merger as described in Section \ref{IC primo merger} and evolve it using {\tt Griffin}. We follow the evolution through the phases of dynamical friction, binary formation, and stellar hardening. We track the evolution of the relative separation between the BHs and calculate the Keplerian semi-major axis ($a$) and eccentricity ($e$) as a function of time.  
Near the pericentre passages, the orbital parameters (particularly the eccentricity) can be quite noisy and poorly defined.  We remove such passages by discarding points where the kinetic energy of the binary exceeds three times its potential energy. 

After alternating between bound and unbound phases, the binary ultimately stabilises into a bound orbit. We define this moment as the time of binary formation ($t_{\rm{b}}$). 

We stop the {\tt Griffin} simulations when the BHB separation reaches the softening length of the most massive BH (i.e. 1 pc), and we manually merge the BHs, placing a single black hole with mass $M=M_{\rm{BH},1}+M_{\rm{BH},2}$ in the centre of mass of the binary and assigning it the centre of mass velocity.

To reconstruct the final stages of the binary evolution (that cannot be reliably simulated with {\tt Griffin}), we use a semi-analytical model (SAM). This model incorporates both the effects of stellar hardening (subscript $\star$) and gravitational wave emission (subscript GW), evolving the semi-major axis of the binary and its eccentricity using the following equations:

\begin{align}
    \frac{da}{dt} &= \left. \frac{da}{dt} \right|_\star + \left.\frac{da}{dt}\right|_{\rm GW}\label{dadt}\\
    \frac{de}{dt} &= \left.\frac{de}{dt}\right|_\star + \left.\frac{de}{dt}\right|_{\rm GW}\label{dedt}\,.
\end{align}
The rate of change of the orbital parameters due to stellar hardening is given by 
\begin{align}
    \left. \frac{da}{dt} \right|_\star  = -a^2\frac{HG\rho}{\sigma}\label{da/dt_star}\\ \left. \frac{de}{dt} \right|_\star = a \frac{HKG\rho}{\sigma}\label{de/dt_star}
\end{align}
with $\rho$ and $\sigma$ being, respectively, the stellar density and velocity dispersion within the radius of influence of the binary, and $H$ and $K$ the dimensionless hardening rate and eccentricity growth rate \citep{1996NewA....1...35Q}.  These depend on the binary's mass ratio, eccentricity and separation and can be derived through three-body scattering experiments of the ejection of background stars by the BHB. Tabulated values given in \citet{2006ApJ...651..392S} can be interpolated as required for specific merger configurations.

The rate of change due to GW emission, on the other hand, is obtained from Peters' equations \citep{1964PhRv..136.1224P}:
\begin{align}
    \left. \frac{da}{dt} \right|_{\rm GW}  &= -\frac{64 G^3}{5c^5}\frac{M_1M_2M}{a^3(1-e^2)^{7/2}}\left(1+\frac{73}{24}e^2+\frac{37}{96}e^4\right)\label{dadt_gw} \\
    \left. \frac{de}{dt} \right|_{\rm GW} &=  -\frac{304 G^3}{15c^5}\frac{M_1M_2M}{a^4(1-e^2)^{5/2}}\left(e+\frac{121}{304}e^3\right)\label{dedt_gw}\,,
\end{align}
where $G$ is the gravitational constant, $c$ is the speed of light in vacuum, $M_1$ and $M_2$ are the primary and secondary BH masses, respectively, and $M$ is the total binary mass. 

The initial conditions of the SAM (namely the initial values of $a$, $\rho$ and $\sigma$) are drawn directly from {\tt Griffin} data: we select a snapshot from the simulation at a time $t_{0, \rm{SAM}}$ that satisfies the following criteria: (i) $t_{0, \rm{SAM}}> t_{\rm{b}}$, (ii) at $t_{0, \rm{SAM}}$ the mutual separation between the BHs is decreasing slowly (indicating that the binary system is already "hard"), (iii) we have sufficient snapshots at $t>t_{0, \rm{SAM}}$ to compare the evolution predicted by the SAM with the final part of the evolution observed in {\tt Griffin}. The semi-analytical model is integrated using the Euler method, until one of the following criteria is met: (i) the binary merges or (ii) $t>10^{10}\, \rm yr$. 

Our semi-analytical model admits three free parameters: the initial eccentricity\footnote{We opted to keep $e_0$ as a free parameter, contrary to what we did for $a_0$, which is instead fixed as an initial condition. This choice is made because the scatter on the eccentricity evolution is usually much larger than the one on the semi-major axis, making it challenging to properly select an initial eccentricity value for the integration.} ($e_{0,\rm{SAM}}$), $\alpha$ and $\beta$, where the last two are normalisation parameters that can adjust the value of $H$ and $K$, respectively, to account for more or less efficient stellar hardening, compared to the one obtained from scattering experiments. We then employ a Markov Chain Monte Carlo (MCMC) sampler to compare the semi-analytical prediction with the evolution of $a$ and $e$ computed from {\tt Griffin} data. We use uniform prior distributions and a Gaussian likelihood, defined as the root mean square of our data points with respect to the semi-analytical model. After computing the posteriors, the peak of the likelihood indicates the parameter values that best fit our data points. The time at which the SAM (run with the best-fit parameters) stops is defined as the time of binary coalescence.

To assess the uncertainty on the predicted coalescence time, we randomly extract 100 samples from the posterior distributions of the parameters. We then use these samples to predict the corresponding coalescence times, which provide an estimate of the error in our predictions.
When simulating the second and third merger of the trees, we apply the same methodology, the only difference being in the initial conditions settings (see Sec. \ref{IC following mergers}).

\section{Results and discussion} \label{Results}
\subsection{Evolutionary time scales } \label{Sec. Time scales}

\begin{figure*}[h!]
    \centering
    \includegraphics[width=1\linewidth]{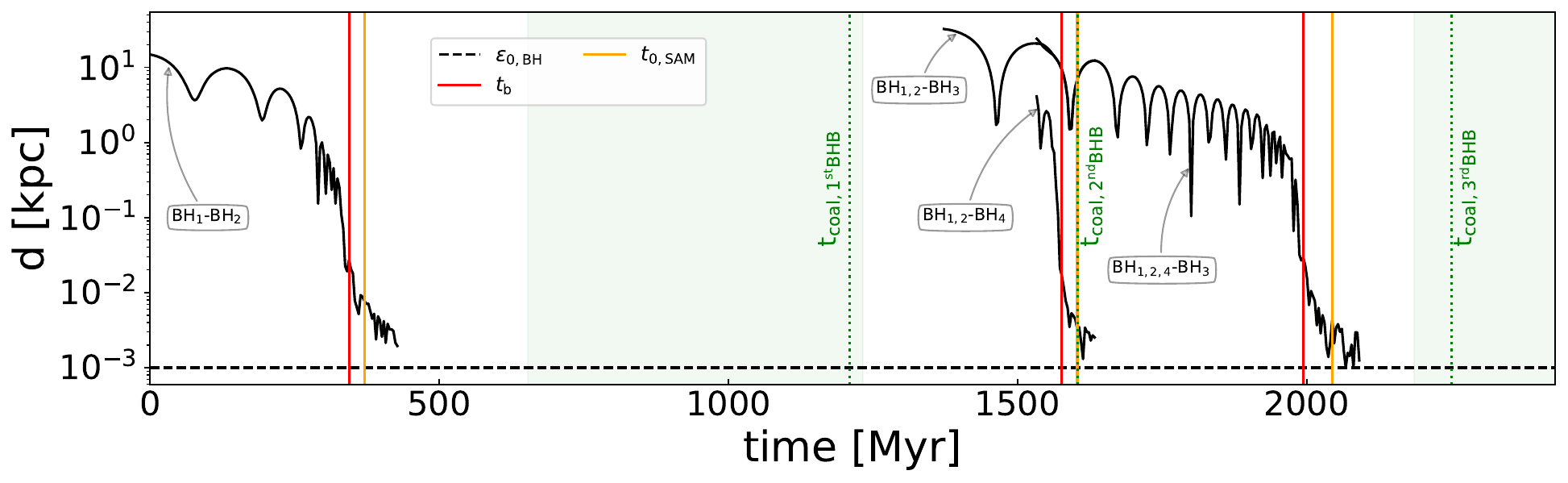}
    \caption{Merger tree 197109 - HR. We show the time evolution of the BHs separation in {\tt Griffin} simulations. The horizontal dashed black line indicates the softening length $e_{0,\rm{BH}}=1$pc.
    Red vertical lines represent binary formation times; yellow vertical lines represent initial times chosen for the SAM; green vertical lines are the predicted coalescence times, with the corresponding uncertainties drawn as green shaded areas. Note that for the second BHB the predicted time of coalescence almost overlaps with $t_{0,\rm{SAM}}$. This is because the eccentricity at binary formation is so extreme that the system merges immediately due to GW emission. As a consequence, this particular system does not form a triplet.}
    \label{fig:evolution of 197109}
\end{figure*}

In Fig. \ref{fig:evolution of 197109} we present the merger tree 197109 - HR as an example of our results, and we plot the BH separation throughout the evolution (analogous figures for the other merger trees can be found in Appendix \ref{Appendix}). 

The first galactic merger begins with eccentricity $e_0=0.745$ at a separation $d \sim 15$ kpc. This merger forms a bound BHB after $\sim 344$ Myr (first red line). The binary is predicted to coalesce roughly $866$ Myr after its formation (first green line), and around $163$ Myr before the onset of the second merger. 

The second and third galactic mergers, on the other hand, are separated by only $\sim 160$ Myr, leading to the co-evolution of 3 different galaxies. The last galaxy is initialised with a smaller separation from the primary galaxy of the previous merger ($\sim 4$ kpc), compared to the $\sim 32$ kpc separation between the two galaxies at the beginning of the second merger\footnote{This depends solely on when each galactic merger is flagged in TNG100-1}. Due to the aforementioned small separation and the high orbital eccentricity ($e_0 = 0.997$) of the last galactic merger, the last BH (BH$_4$) forms a bound binary with the primary BH (BH$_{1,2}$) from the previous galactic encounter. 

This second binary becomes bound $\sim 43$ Myr after the onset of the last galactic merger (second red line), with extremely high eccentricity ($e_{\rm{b}}$= 0.991). During the hardening phase $e$ increases further, reaching a value of 0.999 after roughly 27 Myr (second yellow line). Given the large mass of the BHB, the small separation and the high eccentricity, its evolution is already GW dominated at this point in time. Therefore, in this particular case, instead of using the SAM to infer the coalescence time, we employ the 
%we do not fit the {\tt Griffin} data with the SAM to infer the coalescence time, since the GW emission is not included in {\tt Griffin}, making the data unreliable. Instead, we compute it using the 
following equation:
\begin{equation}
    T_{\rm{GW}}(a_0, e_0) \sim (768/425)T_{\rm{c}}(1-e_0^2)^{7/2} \qquad \text{for} \ e_0 \rightarrow{1}
    \label{eq: tcoal peters}
\end{equation}
where $T_{\rm{c}}= a_0^4/(4\beta_{\rm{e}})$ is the predicted coalescence time for a circular binary, with $\beta_{\rm{e}} = (64/5)(G^3M_1M_2M)/c^5$ \citep{1964PhRv..136.1224P}. The predicted coalescence time is thus $\sim28$ Myr after binary formation (second green line). As a consequence, this particular system does not form a triplet. 

Finally, the remaining two BHs, i.e. the newly merged one (BH$_{1,2,4}$) and the secondary BH coming from the second galactic encounter (BH$_3$), become bound $\sim 364$ Myr after the coalescence of the second binary (third red line), merging around $256$ Myr after binary formation (third green line). 

A summary of the binary formation and coalescence times for all the merger trees is provided in Table \ref{tab:times summary}. Merger trees marked with an "x" in the third BHB column are those that involve only two major mergers in TNG100-1. Some of the high-resolution simulations are particularly slow to run, and have not been completed; these are indicated in Table \ref{tab:times summary} with a "/". We note, however, that their LR counterparts do merge within a Hubble time, giving an indication of their evolutionary time scales. There are two exceptions among the LR runs: 17187-LR and 197109-LR.
The third galactic merger of tree 17187 occurs immediately after the second, and all three galaxies involved have a similar separation, making it the slowest merger tree to simulate, even at low-resolution (see Fig. \ref{fig:plots 17187-LR}). 
In the 197109-LR simulation, on the other hand, the last two BHs (i.e. the remnant resulting from the coalescence of the second binary and the BH coming from the secondary galaxy of the second galactic merger) seem to stall in the LR run (see Fig. \ref{fig:plots 197109-LR}), contrary to the behaviour observed in the HR version of the same tree (see Fig. \ref{fig:evolution of 197109}).
Finally, binaries that are marked with a (T) form before the coalescence of the previous BHB, thereby creating a triple system rather than a binary. For these systems, we do not include a coalescence time, as we plan to follow their evolution using the three-body integrator {\tt Galcode} \citep{2016MNRAS.461.4419B} (F. Fastidio et al., in preparation).

Considering that we have 5 merger trees, 4 of which completed in at least one of the two resolutions, our sample cannot capture the full statistics of merging binaries. However, we can highlight some important observations that may help us understand these results better and improve future studies: (i) starting from an average separation of a few tens of kpc, the dynamical friction phase that causes the BHs to form a binary typically lasts for a few hundred million years, which is generally consistent with previous studies \citep[e.g.][]{2024MNRAS.529.2150A,2024ApJ...974..204K}; (ii) the binary formation times are mostly compatible between the LR and HR runs; (iii) coalescence times tend to be shorter for HR runs, compared with their LR counterparts. This is generally due to a higher eccentricity at binary formation in the HR simulations (see Sec. \ref{Sec: eccentricity}). However, the first binary of 125027 represents an exception in this regard: the predicted coalescence time is shorter in the LR run, compared to the HR one; this is because, while the eccentricity at binary formation is roughly the same for both resolutions, in the LR simulation $e$ increases more during the hardening phase, bringing the binary to coalescence earlier (see Fig. \ref{fig:plots 125027-LR} and \ref{fig:plots 125027-HR}). (iv) Three of the five merger trees form a triple BH system. Tree 168390 forms a triplet in both the high and low resolution runs. Trees 125028 and 197109, on the other hand, form a triplet only in the LR simulations. This outcome has implications for merger time scales and PTA signals, which will be discussed in our forthcoming paper.  

\begin{figure}
    \centering
    \includegraphics[width=1\linewidth]{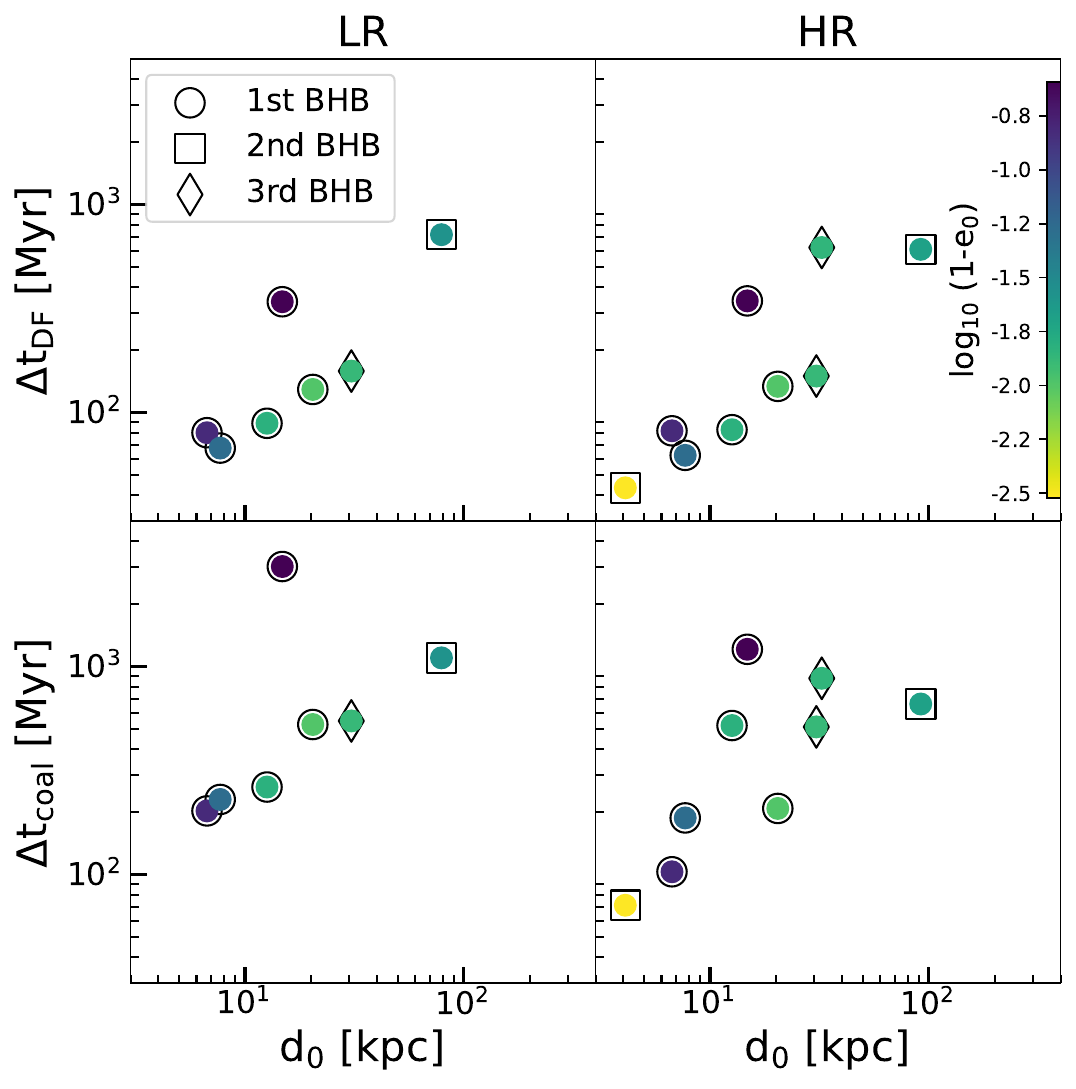}
    \caption{For each BHB formed, we plot the dynamical friction time scale ($\Delta t_{\rm{DF}}=t_{\rm{b}}-t_0$, top panels), and the total coalescence time ($\Delta t_{\rm{coal}}=t_{\rm{coal}}-t_0$, bottom panels) as a function of the initial distance between the progenitor galaxies ($d_0$). Data are colour coded according to the initial eccentricity of the galactic merger. Note that we chose to use log$_{10} (1-e_0)$ to make the colour gradient clearer, so the colour bar goes from high (yellow) to low (blue) eccentricity values. Panels on the left refer to LR runs, while panels on the right show data from their HR counterparts. Empty circles, squares and diamonds represent the first, second and third binary in each tree, respectively.}
    \label{fig:time scales}
\end{figure}

In Fig. \ref{fig:time scales} , we show the dependence of the merger time-scales on three parameters (i) the initial distance between the two progenitor galaxies ($d_0$); (ii) the initial orbital eccentricity of the galactic merger ($e_0$); and (iii) the resolution of the simulation. The four panels illustrate the value of the dynamical friction time scale ($\Delta t_{\rm{DF}}=t_{\rm{b}}-t_0$, top panels), and the total coalescence time ($\Delta t_{\rm{coal}}=t_{\rm{coal}}-t_0$, bottom panels) as a function of the initial separation between the progenitor galaxies. The data points are colour-coded according to the initial eccentricity of the galactic merger.
%\footnote{We use log$_{10}(1-e_0)$ to make the color gradient more visible.}
In the left column, we show data drawn from the LR runs, while on the right we plot data from their HR counterparts. We notice a clear correlation between the initial separation and the dynamical friction time scale. However, for comparable $d_0$ values, a higher eccentricity $e_0$ results in a shorter $\Delta t_{\rm{DF}}$. There is, however, an exception to these general trends: in the upper left panel, we can see two points with roughly the same values of $d_0$ and $e_0$ (the two green diamonds), but with significantly different values of $\Delta t_{\rm{DF}}$. These data points refer to the third BHB in the mergers 197109 (high $\Delta t_{\rm{DF}}$ value) and 125028 (low $\Delta t_{\rm{DF}}$ value). This inconsistency is likely due to the differences in masses and mass ratios of the two mergers, including both the black hole, stellar, and dark matter components. The second system is generally more massive and has a higher mass ratio among all components of the primary and secondary progenitors, which accelerates the dynamical evolution. Finally, we note that the resolution of the simulations does not appear to affect $\Delta t_{\rm{DF}}$, as the values of $t_{\rm{b}}$ remain largely consistent between the LR and HR runs, as previously mentioned.

The bottom panels depict a similar scenario regarding $\Delta t_{\rm{coal}}$. 
Here, a higher $e_0$ value does not necessarily lead to a shorter $\Delta t_{\rm{coal}}$. This is because there is not a perfect correlation between $e_0$ and $e_{\rm{b}}$ (see Sec. \ref{Sec: eccentricity}) and the final part of the binary evolution (i.e. hardening and GW emission phases) can be much faster if $e_{\rm{b}}$ is high. Moreover, there is a difference between LR and HR runs, where coalescence times are on average shorter for high-resolution simulations. 

Lastly, we do not find any specific dependence of these time scales on the sequential nature of these mergers, with no significant differences observed between the first, second, and third black hole binaries formed (indicated by empty circles, squares, and diamonds, respectively). A longer evolutionary timescale might be expected in cored stellar profiles due to less efficient dynamical friction and hardening in a lower density environment. However, our chosen profiles for the second and third merger galaxies may not be sufficiently shallow to show this effect.

\begin{table*}
    \centering
    \begin{tabular}{c|ccc|ccc|ccc}
         Tree's ID&   \multicolumn{3}{c|}{1st BHB}&   \multicolumn{3}{c|}{2nd BHB}&   \multicolumn{3}{c}{3rd BHB}\\
         \hline
         &   $t_0$&t$_{\rm{b}}$ [Myr]&  t$_{\rm{coal}}$ [Myr]&   $t_0$&t$_{\rm{b}}$ [Myr]&  t$_{\rm{coal}}$ [Myr]&   $t_0$&t$_{\rm{b}}$ [Myr]& t$_{\rm{coal}}$ [Myr]\\
         \hline
         17187-LR&   0&129.0&  526.1$^{+36.5}_{-35.0}$&   /&/&  /&   /&/& /\\
         17187-HR&   0&133.5&  207.7$^{+2.9}_{-5.0}$&   /&/&  /&   /&/& /\\
         \hline
         125027-LR&   0&88.7&  263.3$^{+22.7}_{-19.7}$&   752.8&1468.8&  1853.2 $^{+200.6}_{-105.8}$&   x&x& x\\
         125027-HR&   0&82.6&  520.2$^{+135.1}_{-104.9}$&   /&/&  /&   x&x& x\\
         \hline
         125028-LR&   465.8&545.7&  667.9$^{+8.4}_{-54.9}$&   0&615.4$^{\rm{(T)}}$&  (T)&   1691.0&1849.1& 2238.4$^{+15.3}_{-20.7}$\\
         125028-HR&   465.8&547.4&  568.8$^{(1)}$&   0&608.6&  660.3$^{(1)}$&   1691.0&1840.6& 2203.4$^{(1)}$\\
         \hline
         168390-LR&   314.9&382.3&  544.3$^{+233.2}_{-76.6}$&   0&423.9$^{\rm{(T)}}$&  (T)&   x&x& x\\
         168390-HR&   314.9&377.1&  501.9$^{+38.7}_{-18.8}$&   0&420.4$^{\rm{(T)}}$&  (T)&   x&x& x\\
         \hline
 197109-LR&  0&340.7& 3025.9$^{+548.4}_{-1889.3}$&  1533.1&1703.4$^{\rm{(T)}}$& (T)&  /&/&/\\
 197109-HR&  0&344.0& 1210.1$^{+23.2}_{-556.4}$&  1533.1&1576.5& 1604.2$^{(1)}$&  1372.7&1994.0&2250.2$^{+179.3}_{-64.4}$\\
    \end{tabular}
    \caption{We report key times in our simulations: (i) initial time of the galactic mergers ($t_0$); (ii) times of binary formation ($t_{\rm{b}}$); (iii) coalescence times ($t_{\rm{coal}}$). Here $t_0=0$ is the time at which the first galactic merger of each tree is initialised. The cosmological time of the first merger of each tree can be obtained from redshift values shown in Table \ref{tab:mergers params}.  We report $t_0$, $t_{\rm{b}}$ and $t_{\rm{coal}}$ of each BHB formed in the simulated merger trees, both at low and high resolution (LR and HR). The "/" sign denotes simulations that have not finished running; the "x" below the "3rd BHB" column refer to trees that contain only 2 mergers; (T) means that the system forms a triplet. Finally, $^{(1)}$ indicates that $t_{\rm{coal}}$ has been computed using Eq. \ref{eq: tcoal peters}. }
    \label{tab:times summary}
\end{table*}

\subsection{Eccentricity} \label{Sec: eccentricity}
One of the key parameters for determining merger timescales of BHBs is their eccentricity. Unfortunately, this parameter is also the most sensitive to stochastic effects caused by resolution \citep{2020MNRAS.497..739N, 2023MNRAS.526.2688R, 2024MNRAS.532..295F}. In this section, we will discuss the impact of eccentricity on our results and how it is affected by the resolution we have adopted.

\begin{figure}
    \centering
    \includegraphics[width=1\linewidth]{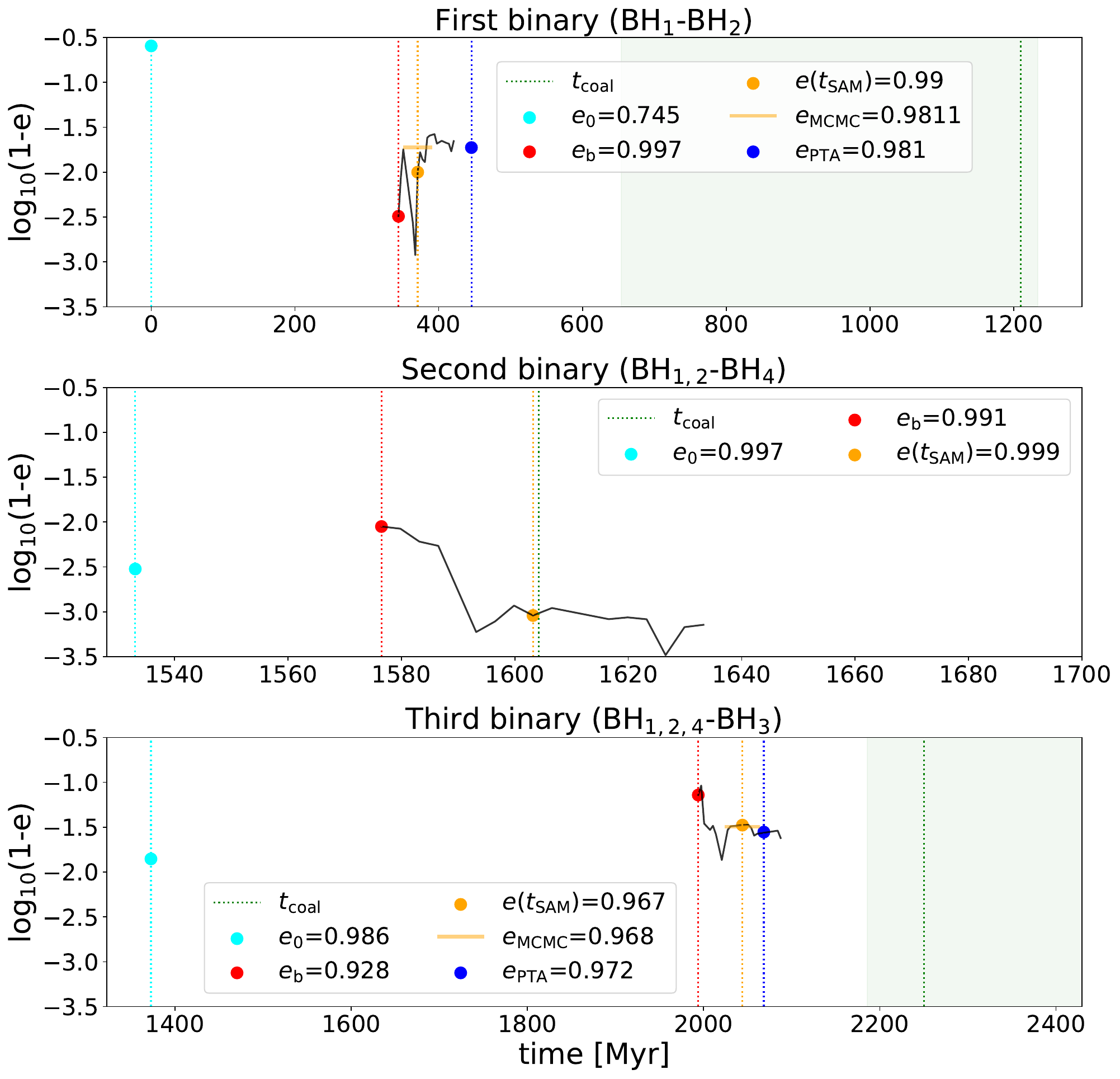}
    \caption{Evolution of the orbital eccentricity of the first (top panel), second (middle panel) and third (bottom panel) BHB formed in Tree 197109-HR.  We highlight with coloured vertical lines the key times in the binary's evolution: initial time of the galactic merger ($t_0$, cyan),  time of binary formation ($t_{\rm{b}}$, red), time where we start the SAM ($t_{\rm{SAM}}$, orange), time when the binary enters the PTA band ($t_{\rm{PTA}}$, blue) and time of coalescence ($t_{\rm{coal}}$, green). For each time, we also report the corresponding eccentricity value,  plotted as a point, following the same colour scheme. The horizontal orange line denotes the MCMC best-fit value of the initial eccentricity used for the semi-analytical evolution.}
    \label{fig:ecc 197109-2M}
\end{figure}

Fig. \ref{fig:ecc 197109-2M} shows the evolution of the Keplerian eccentricity of the three binaries formed in 197109-HR (the same simulation of Fig. \ref{fig:evolution of 197109}). The first panel presents data referring to the first BHB. The galactic merger starts with a moderately high eccentricity ($e_0=0.745$, cyan dot), while the resulting BHB forms with $e_{\rm{b}}=0.997$ (red dot). During the hardening phase, $e$ decreases slightly (at $t=370.7$ Myr $e=0.990$, orange dot), although the data show significant noise. This is why we opted to keep the initial eccentricity of our SAM as a free parameter, fitting it using an MCMC method, as mentioned in Sec. \ref{sec:time evol}. The best fit value we find is  $e_{\rm{MCMC}}=0.981$, which is indeed somewhat lower.

Similarly, the second panel of Fig. \ref{fig:ecc 197109-2M} displays the evolution of the eccentricity of the second binary over time. It is important to note that this binary forms as a result of the third galactic merger. In this case, the primary BH is formed from the merger of the first binary, while the secondary BH is brought in from a fourth galaxy. The third galactic merger starts with $e_0=0.997$, and the resulting BHB has a slightly lower but still significant eccentricity ($e_{\rm{b}}=0.991$). After binary formation, the eccentricity approaches 1, reaching $e=0.999$ at $t=1.60$ Gyr. This high value means we cannot ignore GW emission, which drives the binary to coalescence. 

Finally, the third panel shows data referring to the third BHB, formed when the remnant of the second binary and the BH from the secondary galaxy of the second galactic merger become bound. The second galactic merger starts with $e_0=0.986$, while the formed binary has $e_{\rm{b}}=0.928$, slightly lower than the original. Subsequently, $e$ increases again, reaching $e=0.967$ at $t=2.04$ Gyr. This value is consistent with what we obtained from the MCMC fit. 
Analogous plots for the other merger trees can be found in Appendix \ref{Appendix}. 

A summary of the eccentricity values at key times is presented in Table \ref{tab: eccentricity}. We also report the SAM-predicted eccentricity upon entry in the PTA band ($e_{\rm{PTA}}$). As the reference for lowest frequency observable by PTA, we use $f_{\rm{GW,p}}=10^{-9}$ Hz, where $f_{\rm{GW,p}}$ is the peak gravitational wave frequency, computed via:

\begin{equation}
    f_{\rm{GW,p}} = \frac{1}{\pi}\sqrt{\frac{GM}{[a(1-e^2)]^3}}(1+e)^{1.1954}
    \label{eq: f GW peak}
\end{equation}
where $M$ is the total mass of the binary \citep{2003ApJ...598..419W}. 

These values are provided only for binaries for which the data permit SAM predictions. This includes binaries that do not form triple BH systems and those that are not evolved using Eq. \ref{eq: tcoal peters}. To define $e_{\rm{PTA}}$, we analyse the time evolution of the semi-major axis and orbital eccentricity obtained from the semi-analytical model. These parameters are then used to calculate $f_{\rm{GW,p}}$. 

\begin{figure}
    \centering
    \includegraphics[width=1\linewidth]{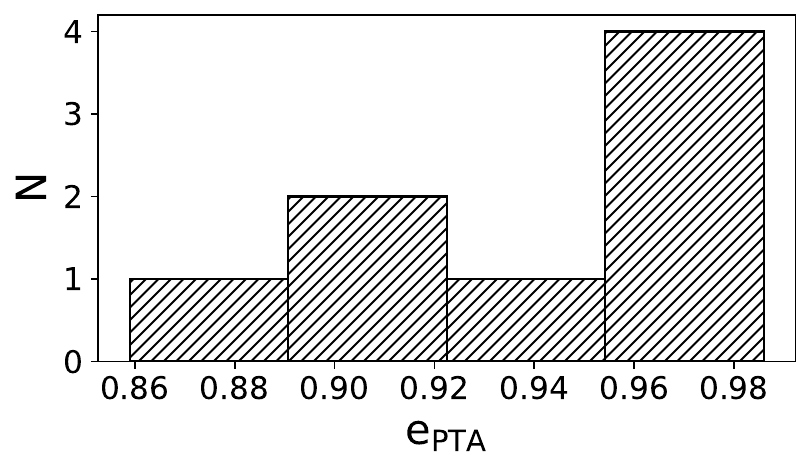}
    \caption{Histogram of the eccentricity upon entry in the PTA band of all binaries for which we can compute the peak gravitational wave frequency ($f_{\rm{GW,p}}$).}
    \label{fig:histo ecc PTA}
\end{figure}

All the 8 binaries for which we can compute $f_{\rm{GW,p}}$ enter the PTA band with an orbital eccentricity $e>0.85$, as shown in Fig. \ref{fig:histo ecc PTA}. This reinforces the importance of incorporating the eccentricity of binaries when analysing PTA data. A population of eccentric binaries is expected to flatten the GWB signal at lower frequencies while boosting it at higher frequencies \citep[e.g.][]{2015ASSP...40..147S}. This is caused by the emission at harmonics higher than twice the orbital frequency. Moreover, eccentric systems evolve faster, which may lead to an overall attenuation of the GWB across all frequencies \citep[e.g.][]{2017MNRAS.470.1738C, 2017MNRAS.471.4508K}. Additionally, close pericentre passages may result in burst-like GW emissions that could be important when interpreting PTA data.

\begin{table*}
    \centering
    \begin{tabular}{c|cccc|cccc|cccc}
         Tree's ID&  \multicolumn{4}{c|}{1st BHB}&  \multicolumn{4}{c|}{2nd BHB}&  \multicolumn{4}{c}{3rd BHB}\\
         \hline
         &  $e_0$&$e_{\rm{b}}$&  $e_{\rm{MCMC}}$ &e$_{\rm{PTA}}$&  $e_0$&$e_{\rm{b}}$&  $e_{\rm{MCMC}}$ &e$_{\rm{PTA}}$&  $e_0$&$e_{\rm{b}}$& $e_{\rm{MCMC}}$ &e$_{\rm{PTA}}$\\
         \hline
         17187-LR&  0.990&0.744&  0.721 &0.905&  / &/&  / &/&  / &/& / &/\\
         17187-HR&  0.990&0.930&  0.985 &0.986&  / &/&  / &/&  / &/& / &/\\
         \hline
         125027-LR&  0.985&0.467&  0.643 &0.930&  0.974&0.871&  0.918 &0.971&  x &x& x &x\\
         125027-HR&  0.985&0.510&  0.700 &0.859&  /&/&  / &/&  x &x& x &x\\
         \hline
         125028-LR&  0.850&0.982&  0.986 &(T)&  0.980$^{\rm{(T)}}$&0.940$^{\rm{(T)}}$&  (T)&(T)&  0.987&0.571& 0.600 &0.899\\
         125028-HR&  0.850&0.998&  0.998$^{(1)}$ &$^{(1)}$ &  0.980&0.995&  0.997$^{(1)}$&$^{(1)}$&  0.987&0.969& 0.986$^{(1)}$&$^{(1)}$ \\
         \hline
         168390-LR&  0.947&0.794&  0.883 &(T)&  0.775$^{\rm{(T)}}$&0.952$^{\rm{(T)}}$&  (T)&(T)&  x &x& x &x\\
         168390-HR&  0.947&0.770&  0.528 &(T)&  0.775$^{\rm{(T)}}$&0.632$^{\rm{(T)}}$&  (T)&(T)&  x &x& x &x\\
         \hline
 197109-LR& 0.745&0.486& 0.703 &(T)& 0.997$^{\rm{(T)}}$&0.367$^{\rm{(T)}}$& (T)&(T)& /&/&/ &/\\
 197109-HR& 0.745&0.997& 0.981 &0.981& 0.997&0.991& 0.999$^{(1)}$ &$^{(1)}$ & 0.986&0.928&0.968 &0.972\\
    \end{tabular}
    \caption{Summary of eccentricity values for all binaries at relevant times: $e_0$ is the initial eccentricity of the galactic merger; $e_{\rm{b}}$ is the eccentricity at binary formation; $e_{\rm{MCMC}}$ is the best fit value for the initial eccentricity used in the SAM (obtained from the MCMC); $e_{\rm{PTA}}$ is the eccentricity with which the binary enters the PTA band (i.e. when $f_{\rm{GW,p}}=10^{-9}$ Hz). The symbols "/", "x", (T) and $^{(1)}$ have the same meaning as in Table \ref{tab:times summary}.}
    \label{tab: eccentricity}
\end{table*}

Following up on the results presented in \cite{2024MNRAS.532..295F}, we investigate the correlation between the initial eccentricity of the galactic merger ($e_0$) and the eccentricity at binary formation ($e_{\rm{b}}$). We expect to find evidence of this correlation up to $e_0\sim0.9$ \citep[see also][]{2022MNRAS.511.4753G}, while for larger values of $e_0$ we anticipate a greater scatter in the results ($\sigma_{\rm{e}_{\rm{b}}}$) which could lead to smaller $e_{\rm{b}}$ values. According to \cite{2020MNRAS.497..739N}, the stochastic effects responsible for the large $\sigma_{\rm{e}_{\rm{b}}}$ decrease as $\sqrt{N}$, where $N$ represents the number of particles used in the simulation. \cite{2023MNRAS.526.2688R}, on the other hand, argue that this behaviour does not hold at extreme initial eccentricities, where stochasticity becomes nearly inevitable and only weakly depends on resolution. As mentioned, in this study we have a limited number of runs, thus our goal is not to establish how standard deviation $\sigma_{\rm{e}_{\rm{b}}}$ decreases (or not) with $N$ (Gualandris et al., in preparation). Nevertheless, our HR results seem to favour a correlation between $e_0$ and $e_{\rm{b}}$ even for high eccentricities, with $\sigma_{\rm{e}_{\rm{b}}}=0.13$ with respect to the 1:1 correlation. On the other hand, in the LR runs, when the initial eccentricities are extremely high, $e_{\rm{b}}$ tends to be lower and $\sigma_{\rm{e}_{\rm{b}}}=0.30$.  

\begin{figure}
    \centering
    \includegraphics[width=1\linewidth]{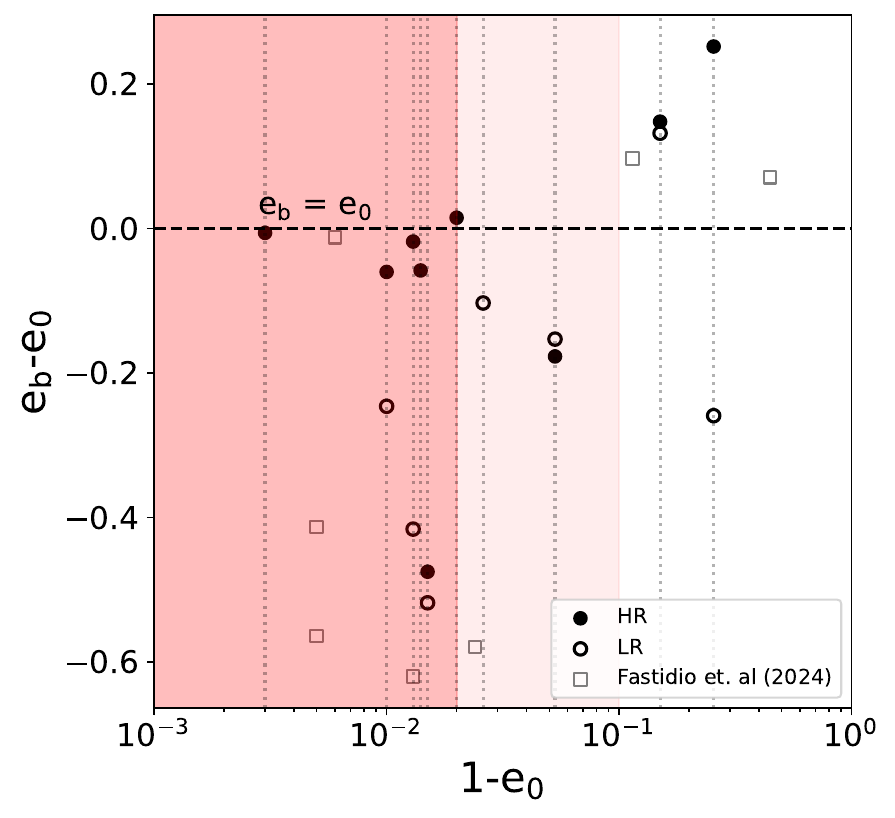}
    \caption{Difference between the eccentricity at binary formation ($e_{\rm{b}}$) and the initial eccentricity of the merger ($e_0$) as a function of the initial galactic merger orbital eccentricity ($e_0$). The dashed horizontal line at $(e_{\rm{b}} - e_0)=0$ highlights where the points should be if the was 1:1 correlation. Empty and filled dots represent data from the LR and HR runs, respectively. The faint red shaded area highlights the region where $e_0>0.9$, while the darker red shaded area defines the region where $e_0>0.98$}
    \label{fig:ecc corr}
\end{figure}

Fig. \ref{fig:ecc corr} shows the difference $(e_{\rm{b}}-e_0)$ as a function of $e_0$, where we include data from all binaries of our simulations. If there was a 1:1 correlation between $e_{\rm{b}}$ and $e_0$, the points should cluster along the horizontal line where $(e_{\rm{b}}-e_0)=0$. Generally, LR runs (empty dots) do not follow this trend: they tend to be more scattered than HR data, and in the region of high initial eccentricities $(e_{\rm{b}}-e_0)$ is always negative. In contrast, the HR runs (filled dots) on average produce binaries with higher $e_{\rm{b}}$, which are more consistent with the original merger eccentricity. For completeness, we also include data from \cite{2024MNRAS.532..295F}\footnote{The eccentricity at binary formation is computed as described in Sec. \ref{sec:methods}, which is slightly different from the method used in \cite{2024MNRAS.532..295F}} (grey squares), which have a resolution comparable with our LR runs, and we generally find a good agreement.

\begin{figure}
    \centering
    \includegraphics[width=1\linewidth]{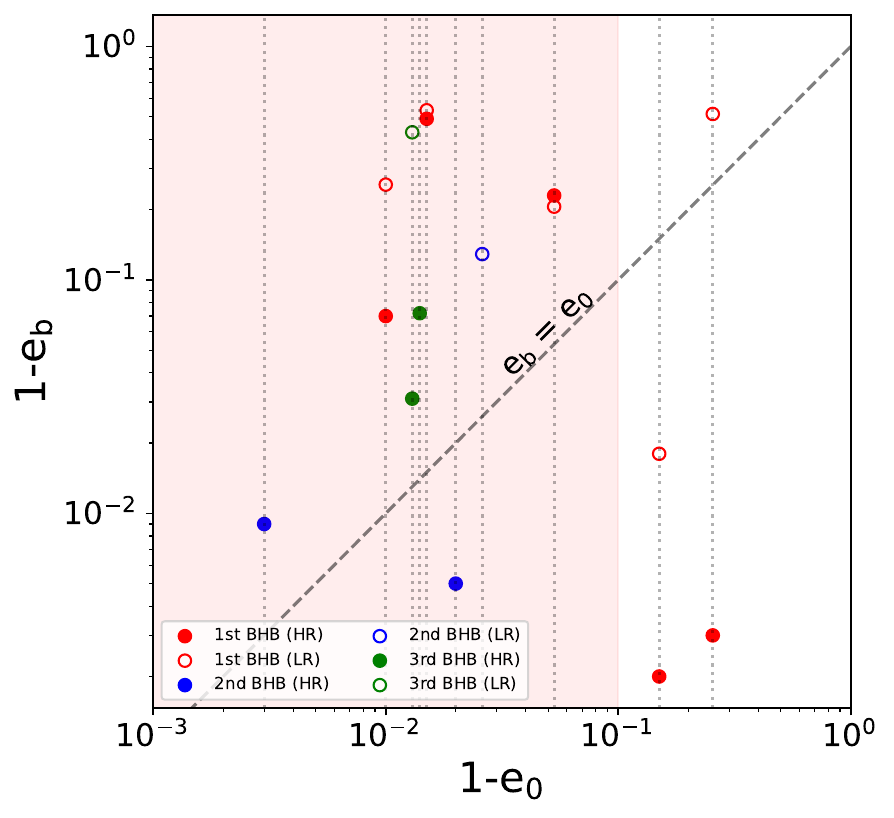}
    \caption{Eccentricity at binary formation ($e_{\rm{b}}$) as a function of the initial galactic merger orbital eccentricity ($e_0$). We use different colours to distinguish between data relative to the first (red), the second (blue) or the third (green) BHB of a tree. Empty and filled dots represent data from the LR and HR runs, respectively. The red shaded area highlights the region where $e_0>0.9$.} 
    \label{fig:ecc corr colors}
\end{figure}

In Fig. \ref{fig:ecc corr colors}, we plot the eccentricity at the time of binary formation as a function of $e_0$. This includes all the binaries from our five merger trees. Different colours are used to distinguish between the first (red), second (blue) and third (green) binary in each tree, while empty and filled dots represent LR and HR runs, respectively. Similar to what is shown in Fig. \ref{fig:ecc corr}, we observe that in the LR runs, when $e_0>0.9$ (red shaded area), the value of $e_{\rm{b}}$ tends to fall below the bisector (dashed line) and the scatter increases. However, there seems to be no significant impact on the correlation due to the sequential nature of these mergers; the first, second, and third binaries exhibit the same trend. In Appendix \ref{Appendix}, Fig. \ref{fig:correlation no log}, we show the same figure in linear scale and and we also include data from our previous work \citep{2024MNRAS.532..295F}, to make the comparison easier.

\section{Conclusions} \label{Conclusions}

We searched for consecutive low-redshift major mergers in the cosmological simulation IllustrisTNG100-1, as these mergers lead to the formation of massive black hole binaries that are observable PTA sources. Our goal was to consistently re-simulate these systems at high resolution using the FMM code {\tt Griffin} down to BH separations of order 1 pc. When initialising the second (or third) galactic merger of the selected merger trees, we accounted for both the effects of core scouring and anisotropy that can result from the preceding galactic merger, as these may influence subsequent mergers. We predicted the final stages of the BHBs' evolution using a semi-analytical model that includes the effects of both stellar hardening and GW emission.  We ran two versions of each simulation, changing their mass resolution, to compare results between the two. Our investigation centred on two primary aspects: the evolutionary timescales of the simulated systems and the evolution of the eccentricity of the formed BHBs. Our main findings are as follows:
\begin{enumerate}
    \item The dynamical friction phase, which leads to BHs forming a binary, typically lasts for a few hundred million years, starting from an average separation of a few tens of kpcs between the progenitor galaxies.
    \item For comparable initial separations of the two galaxies at the onset of the galaxy merger, a higher initial eccentricity $e_0$ of the galaxy merger results in a shorter dynamical friction timescale $\Delta t_{\rm{DF}}$.
    \item Binary formation times are mostly consistent between the LR and HR runs.
    \item In all the finished runs, the BHBs merge within a Hubble time, making them observable PTA sources.
    \item Coalescence times tend to be shorter for HR runs, compared with their LR counterparts. This is typically caused by a higher eccentricity at binary formation in the HR simulations.
    \item The high-resolution runs seem to favour a correlation between $e_0$ and $e_{\rm{b}}$ (i.e. the eccentricity of the galaxy merger and the eccentricity of the binary at its formation) even for high eccentricities, with $\sigma_{\rm{e}_{\rm{b}}}=0.13$ with respect to the 1:1 correlation. On the other hand, the dispersion is larger in their LR counterparts ($\sigma_{\rm{e}_{\rm{b}}}=0.30$), since stochastic effects are stronger when the initial eccentricities are extremely high.
    \item   A larger  $\sigma_{\rm{e}_{\rm{b}}}$ can result in a lower eccentricity at binary formation and thus in longer coalescence times.
    \item All the eight binaries for which we can compute the peak GW frequency ($f_{\rm{GW,p}}$) enter the PTA band with an orbital eccentricity $e>0.85$,  indicating that eccentricity is a crucial parameter to include in PTA data analysis. This may explain or contribute the flattening of the observed GWB signal.
\end{enumerate}

Finally, three out of our five merger trees form a triple black hole system: one was observed in both HR and LR runs, and two others were found only in the LR simulations. We will present the results related to the 3-body interactions of these systems in a follow-up paper, where we will track their dynamical evolution using the 3-body integrator code {\tt Galcode} to predict the GW emission we can expect from such systems.

\begin{acknowledgements}
AS acknowledges the financial support provided under the European Union's H2020 European Research Council (ERC) Consolidator Grant ``Binary Massive Black Hole Astrophysics'' (B Massive, Grant Agreement: 818691). AG would like to thank the Science and Technology Facilities Council (STFC) for support from grant ST/Y002385/1.
The simulations were run on the Eureka2 HPC cluster at the University of Surrey.
\end{acknowledgements}

\bibliographystyle{aa} % style aa.bst
\bibliography{biblio} % your references Yourfile.bib

\newpage

\begin{appendix}
\section{Additional plots}\label{Appendix}
Here we show analogous plots to Fig. \ref{fig:evolution of 197109} and Fig. \ref{fig:ecc 197109-2M} for all the other merger trees, both in their LR and HR versions and point out peculiarities of each one. The colour scheme and the legends are the same used for  Fig. \ref{fig:evolution of 197109} and Fig. \ref{fig:ecc 197109-2M}, unless otherwise stated. Note that (i) the eccentricity upon entry in the PTA band is plotted only for binaries for which $f_{\rm{GW,p}}$ can be computed (i.e. they do not form triplets and their later evolution is predicted using the SAM model); (ii) when the coalescence time of a binary is computed using Eq. \ref{eq: tcoal peters} we do not plot any uncertainty on it; (iii) in this paper we do not report the coalescence time of triple BH systems.

\begin{figure}
\centering
\begin{subfigure}[a]{0.5\textwidth}
   \includegraphics[width=\textwidth]
{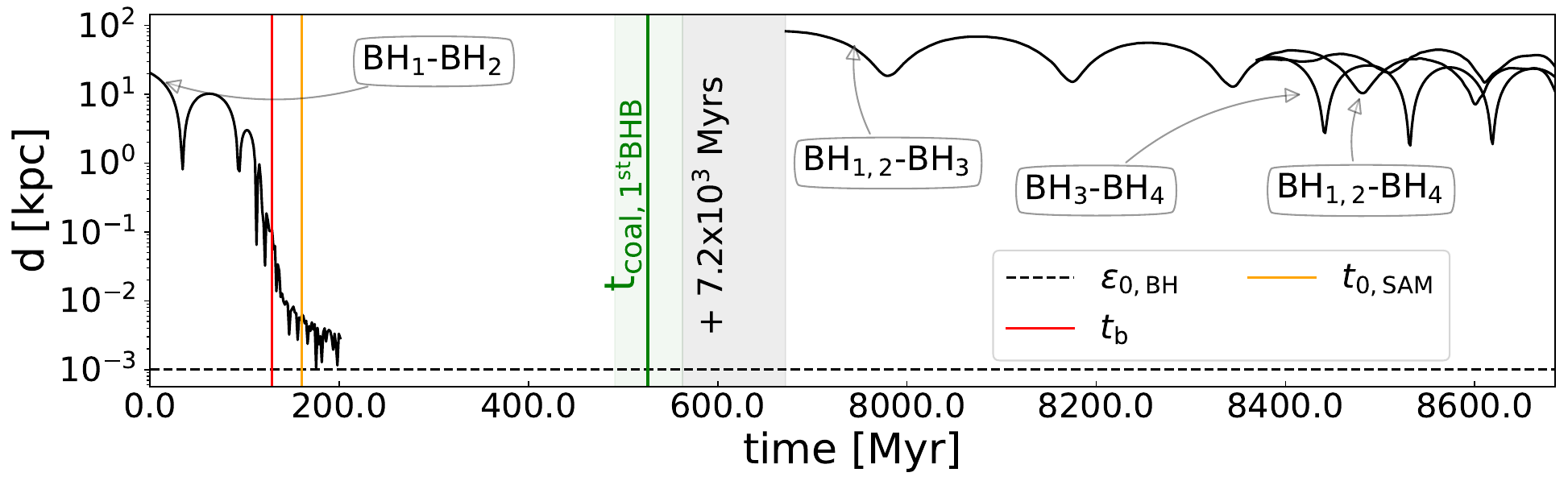}
   \caption{}
   \label{fig:evol 17187-LR}
\end{subfigure}
\hspace{4cm}
   \begin{subfigure}[b]{0.5\textwidth}
   \includegraphics[width=\textwidth]
   {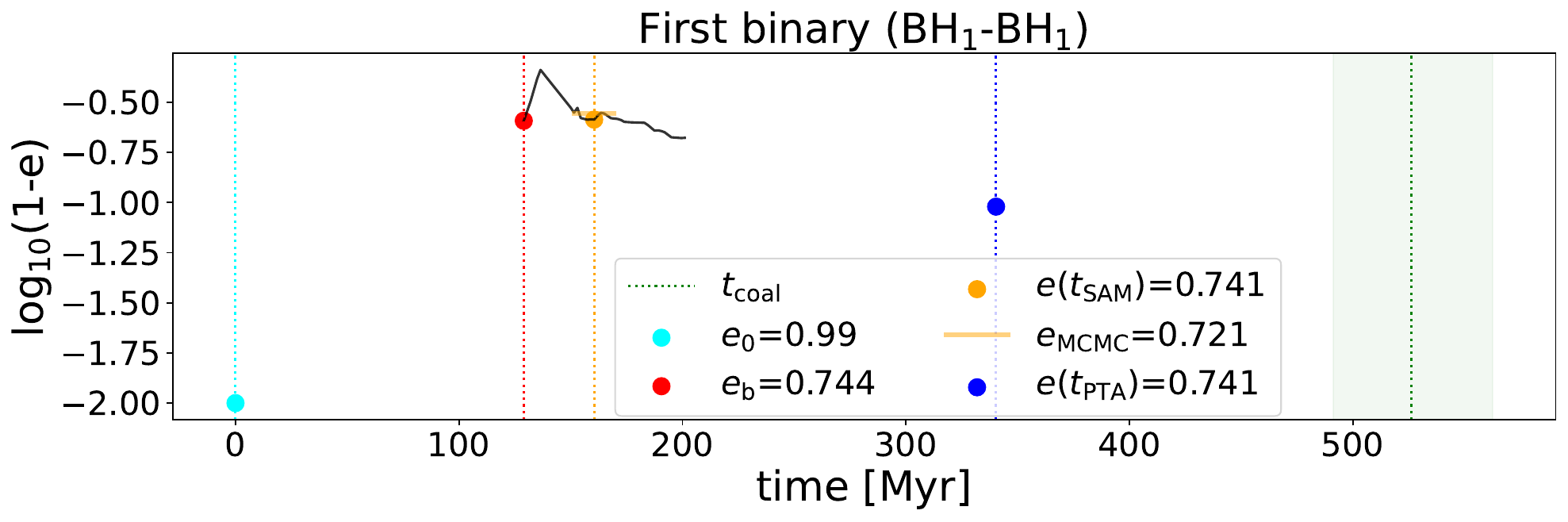}
   \caption{}
   \label{fig:ecc 17187-LR} 
\end{subfigure}
\caption{(a) Merger tree 17187-LR: time evolution of the BHs separation in {\tt Griffin} simulations. The horizontal dashed black line indicates the softening of BHs particles in our simulation. Coloured vertical lines highlight key evolutionary times: initial galactic merger time (cyan), time of binary formation (red), time where we start the semi-analytical model (orange), time of coalescence (green). The green shaded area represents the error on the predicted coalescence time, while the grey area denotes a jump in time of $7.2\times10^3$ Myr, used to make the plot clearer. (b) Evolution of the orbital eccentricity of the first BHB (BH$_1$ and BH$_2$) formed in Tree 17187-LR. Vertical lines are coloured according to the same colour scheme as in panel (a). The blue vertical line shows when the binary is predicted to enter the PTA band by the SAM. For each key time, we plot the value of the orbital eccentricity with a point of the corresponding colour. The horizontal orange line denotes the MCMC best-fit value of the initial eccentricity used for the semi-analytical evolution.} 
\label{fig:plots 17187-LR}
\end{figure}

\begin{figure}
\centering
\begin{subfigure}[a]{0.5\textwidth}
   \includegraphics[width=\textwidth]
{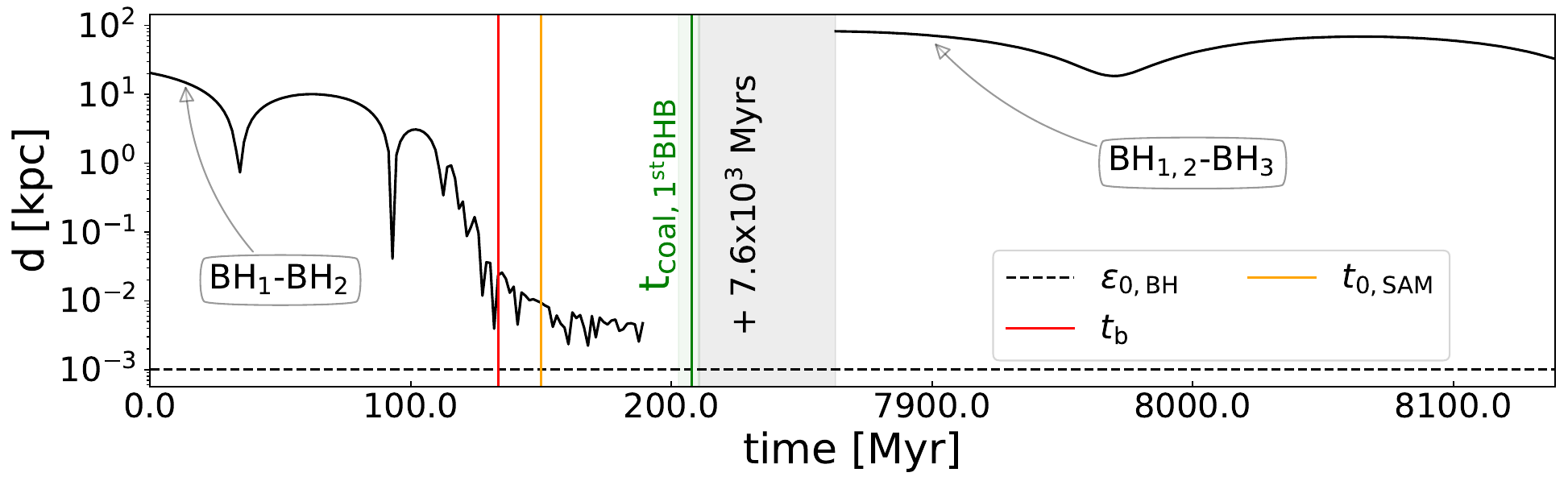}
   \caption{}
   \label{fig:evol 17187-HR}
\end{subfigure}
\hspace{4cm}
   \begin{subfigure}[b]{0.5\textwidth}
   \includegraphics[width=\textwidth]
   {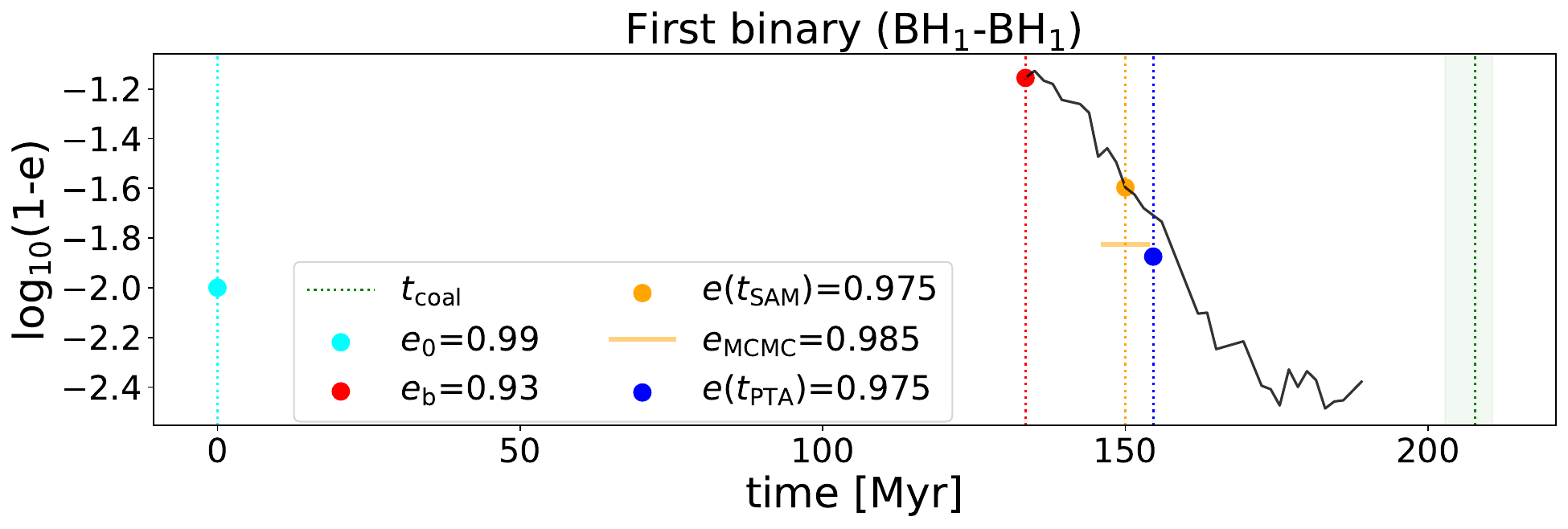}
   \caption{}
   \label{fig:ecc 17187-HR} 
\end{subfigure}
\caption{(a) Merger tree 17187-HR: time evolution of the BHs separation. The plotted grey area denotes a jump in time of $7.6\times10^3$ Myr, used to make the plot clearer. (b) Evolution of the orbital eccentricity of the first BHB (BH$_1$ and BH$_2$) formed in Tree 17187-HR.}
\label{fig:plots 17187-HR}
\end{figure}

Fig. \ref{fig:plots 17187-LR} shows results that refer to the merger tree 17187-LR. In panel (a) one can see that the second and third galactic mergers occur a few Gyrs after the first, both at very low redshifts (z=0.21 and z=0.17). When the third galactic merger is initialised, the three galaxies involved are roughly at the same separation of a few tens of kpcs. This makes the simulation quite slow to form a potential second and third BHB (or triple system). Only data relative to the first BHB eccentricity are thus available and shown in panel (b). Moreover, due to the low redshift of this encounter, there is a high chance that it will not result in a bound BH system within a Hubble time.
In Fig. \ref{fig:plots 17187-HR} we present the same merger tree as Fig. \ref{fig:plots 17187-LR}, but in its high-resolution version. Note that the HR runs are generally slower since more computationally expensive and here it results in a lack of data relative to the third galactic merger. 

\begin{figure}
\centering
\begin{subfigure}[a]{0.5\textwidth}
   \includegraphics[width=\textwidth]
{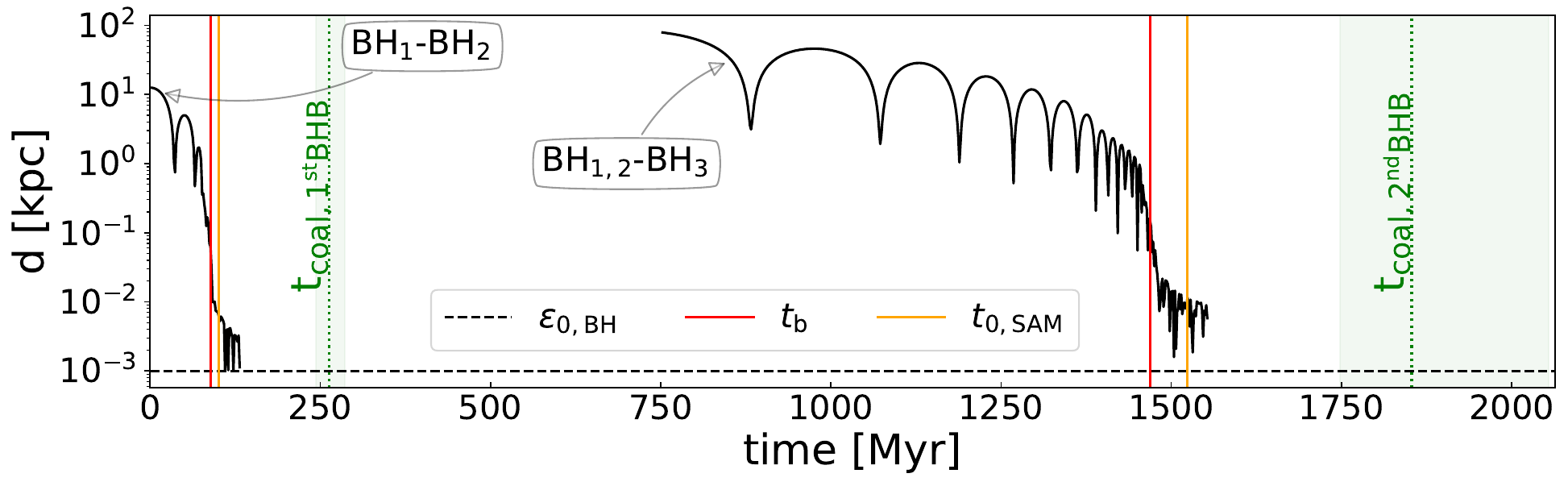}
   \caption{}
   \label{fig:evol 125027-LR}
\end{subfigure}
\hspace{4cm}
   \begin{subfigure}[b]{0.5\textwidth}
   \includegraphics[width=\textwidth]
   {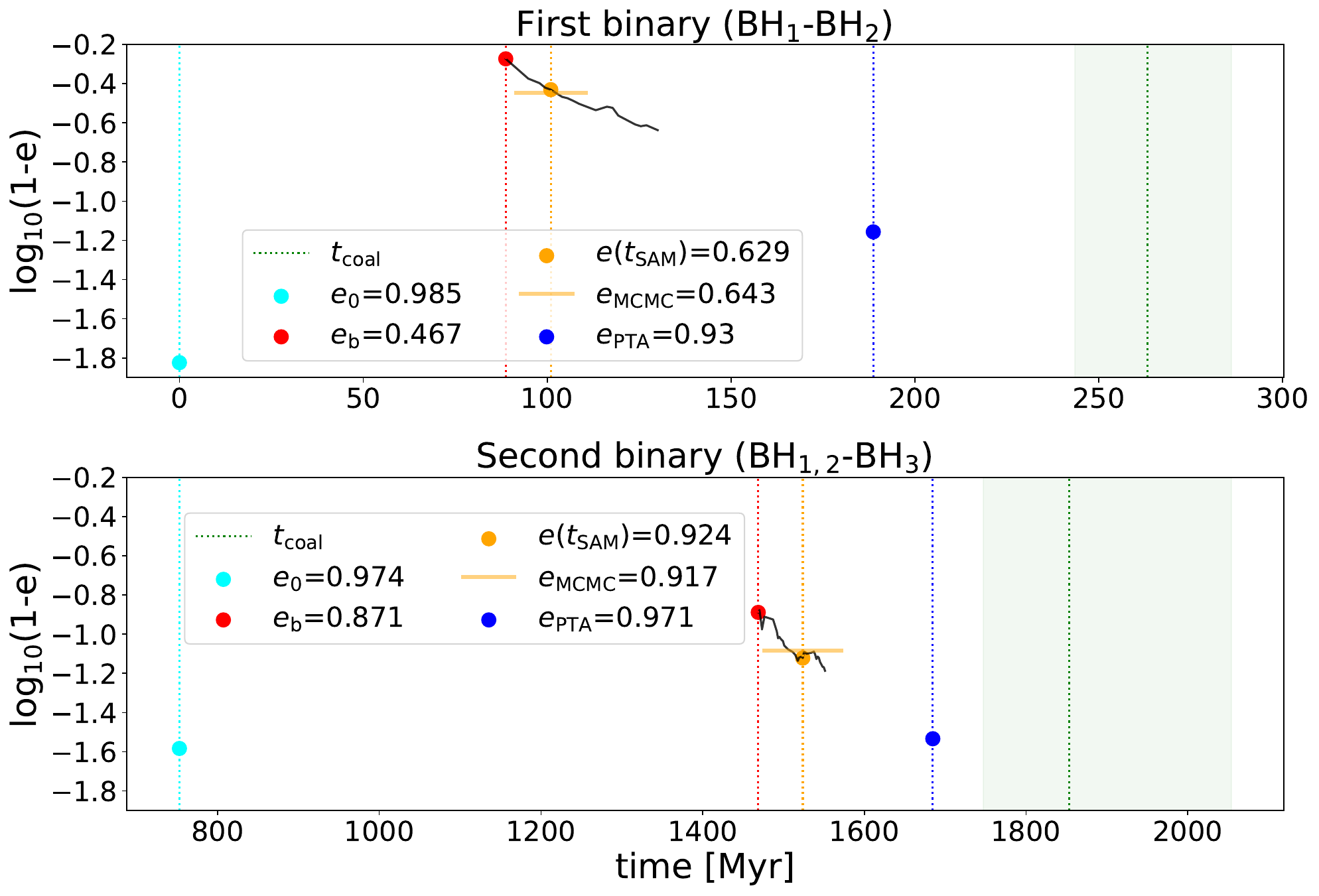}
   \caption{}
   \label{fig:ecc 125027-LR} 
\end{subfigure}
\caption{(a) Merger tree 125027-LR: time evolution of the BHs separation. (b) Evolution of the orbital eccentricity of the first (BH$_1$ and BH$_2$) and second (BH$_{1,2}$ and BH$_3$) binary formed in Tree 125027-LR.}
\label{fig:plots 125027-LR}
\end{figure}

\begin{figure}
\centering
\begin{subfigure}[a]{0.5\textwidth}
   \includegraphics[width=\textwidth]
{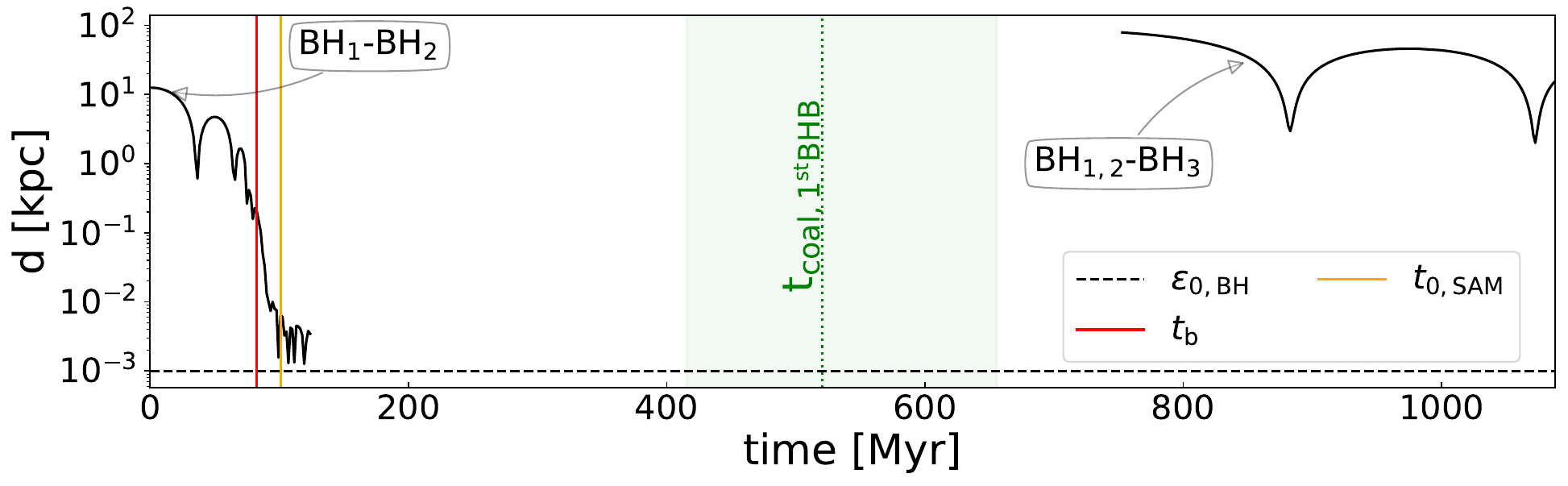}
   \caption{}
   \label{fig:evol 125027-HR}
\end{subfigure}
\hspace{4cm}
   \begin{subfigure}[b]{0.5\textwidth}
   \includegraphics[width=\textwidth]
   {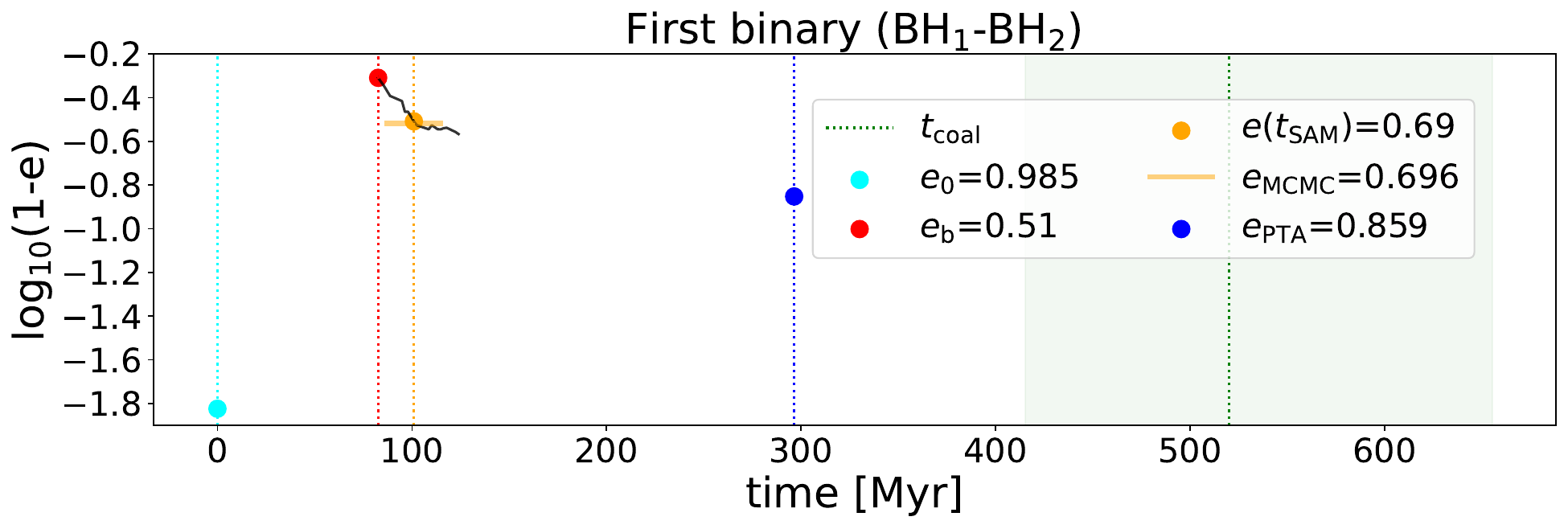}
   \caption{}
   \label{fig:ecc 125027-HR} 
\end{subfigure}
\caption{(a) Merger tree 125027-HR: time evolution of the BHs separation. (b) Evolution of the orbital eccentricity of the first BHB (BH$_1$ and BH$_2$) formed in Tree 125027-HR.}
\label{fig:plots 125027-HR}
\end{figure}

In panel (a) of Fig. \ref{fig:plots 125027-LR}, showing the time evolution of the BHs separation of merger tree 125027-LR, we can see an example of complete 2-merger tree, while the high-resolution version has yet to form the second bound BHB (see Fig. \ref{fig:plots 125027-HR}).

\begin{figure}
\centering
\begin{subfigure}[a]{0.5\textwidth}
   \includegraphics[width=\textwidth]
{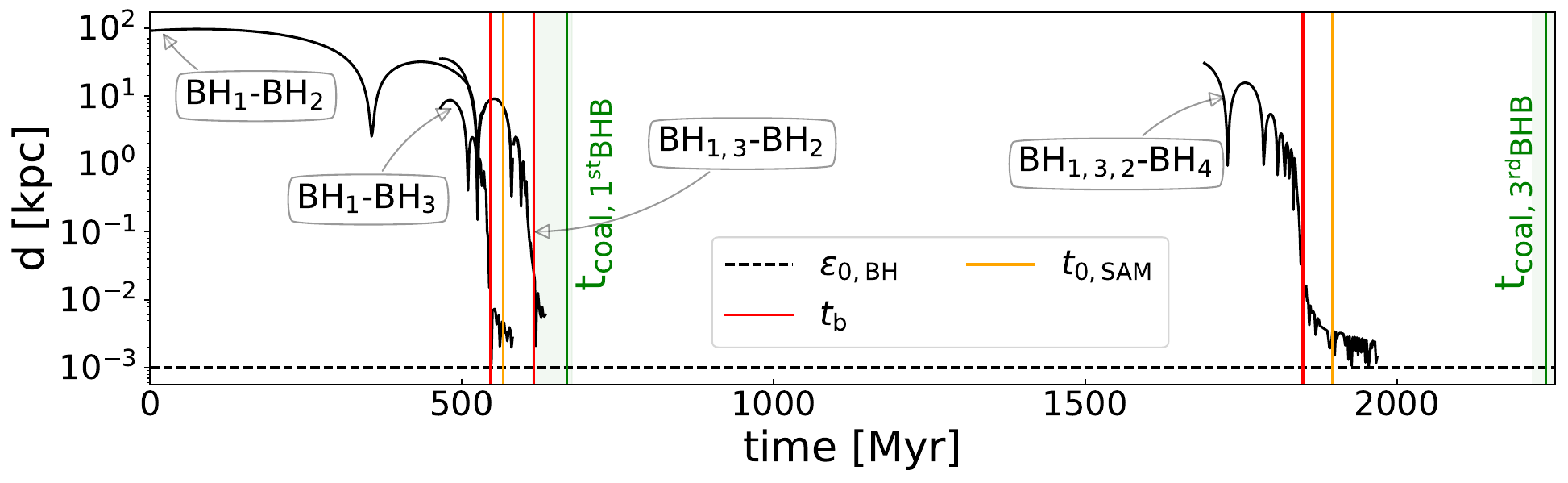}
   \caption{}
   \label{fig:evol 125028-LR}
\end{subfigure}
\hspace{4cm}
   \begin{subfigure}[b]{0.5\textwidth}
   \includegraphics[width=\textwidth]
   {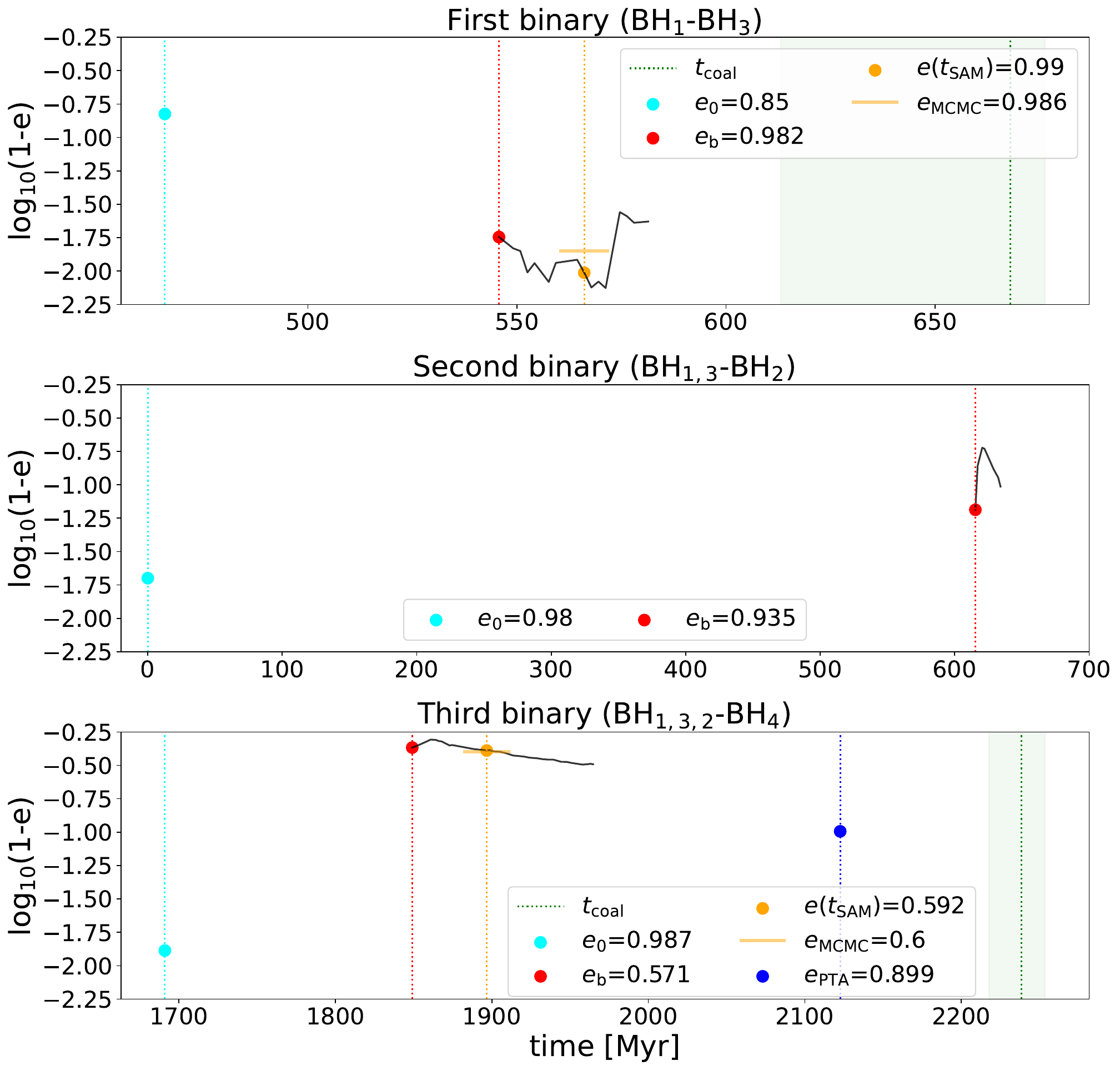}
   \caption{}
   \label{fig:ecc 125028-LR} 
\end{subfigure}
\caption{(a) Merger tree 125028-LR: time evolution of the BHs separation. (b) Evolution of the orbital eccentricity of the first (BH$_1$ and BH$_3$), second (BH$_{1,3}$ and BH$_2$, triplet) and third (BH$_{1,3,2}$ and BH$_4$) binary formed in Tree 125028-LR.}
\label{fig:plots 125028-LR}
\end{figure}

\begin{figure}
\centering
\begin{subfigure}[a]{0.5\textwidth}
   \includegraphics[width=\textwidth]
{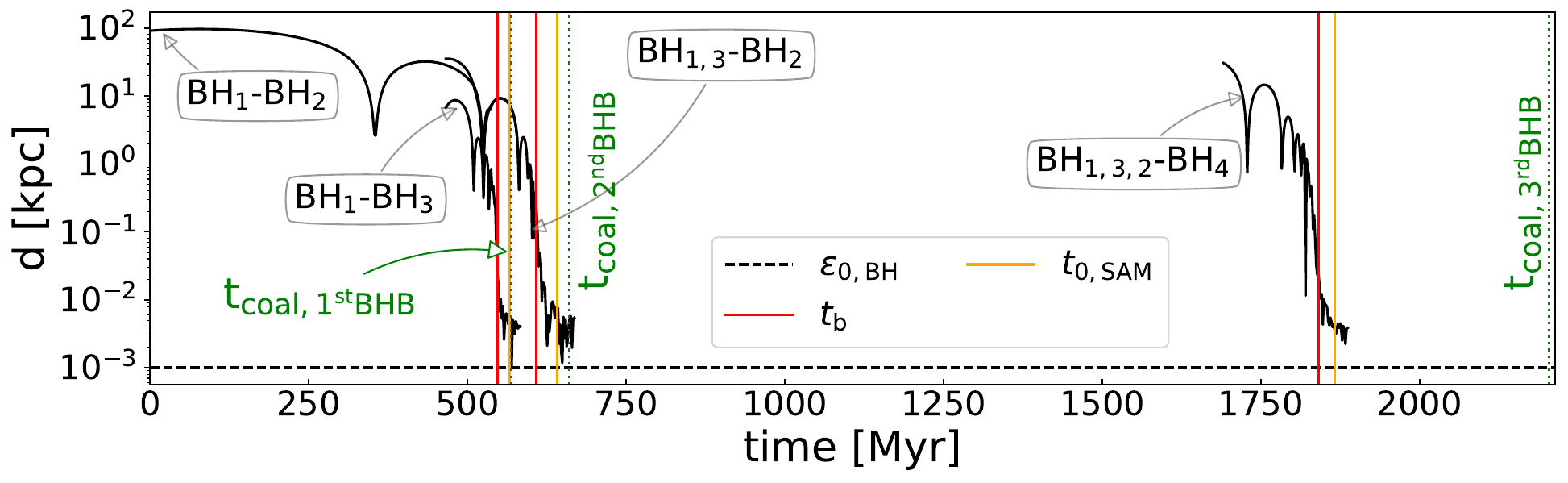}
   \caption{}
   \label{fig:evol 125028-HR}
\end{subfigure}
\hspace{4cm}
   \begin{subfigure}[b]{0.5\textwidth}
   \includegraphics[width=\textwidth]
   {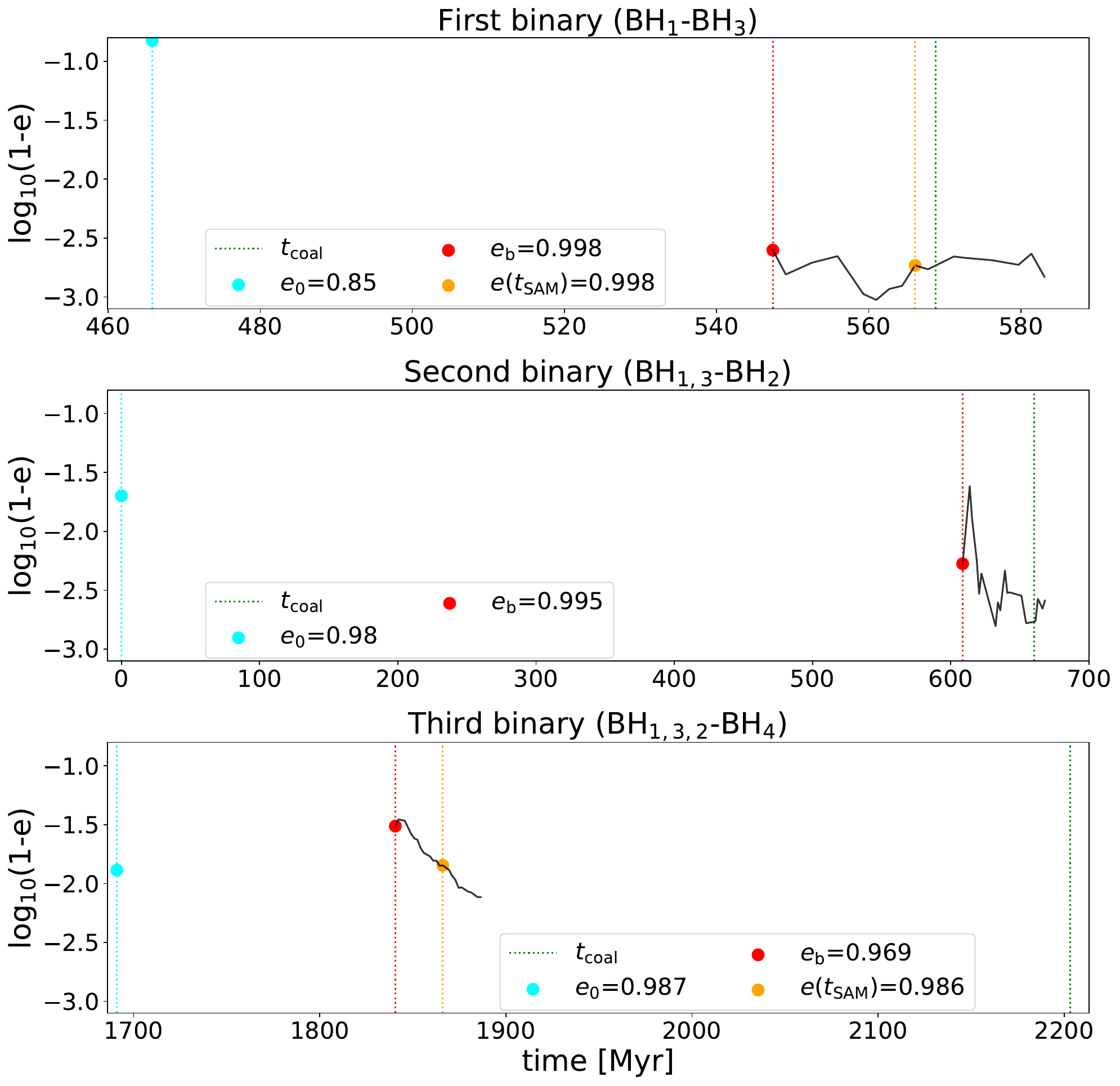}
   \caption{}
   \label{fig:ecc 125028-HR} 
\end{subfigure}
\caption{(a) Merger tree 125028-HR: time evolution of the BHs separation. (b) Evolution of the orbital eccentricity of the first (BH$_{1}$ and BH$_3$), second (BH$_{1,3}$ and BH$_2$) and third (BH$_{1,3,2}$ and BH$_4$) binary formed in Tree 125028-HR.}
\label{fig:plots 125028-HR}
\end{figure}

Fig. \ref{fig:plots 125028-LR} shows one of the complete 3-merger trees that also forms a triple BH system in its low-resolution version (tree 125028-LR). We note that the first BHB forms between BH$_1$ (coming from the primary galaxy of the first galactic merger) and BH$_3$ (from the secondary galaxy of the second galactic merger). This is because (i) the initial separation between galaxy 1 and galaxy 3 is smaller than the separation between galaxy 1 and galaxy 2 and (ii)  BH$_1$ and  BH$_3$ are more massive than  BH$_2$, making the dynamical friction more efficient on them. In panel (a) we can see that the time of coalescence of the first BHB is predicted to be after the binary formation time of the second BHB, thus forming a triple system. We do not plot the initial SAM time and coalescence time of the second BHB, since we will follow the evolution of this triple system with the 3-body integrator {\tt Galcode} and present our results in F. Fastidio et al. (in preparation). When looking at the eccentricity evolution in panel (b), note that: (i) the initial eccentricity $e_0$ plotted in the "First BHB" panel is the eccentricity of the second galactic merger (since the involved BHs are  BH$_1$ and  BH$_3$), (ii) the value of $e_0$ in the "Second BHB" panel is the orbital eccentricity of the first galactic merger (for the same reason). 
The HR version of this tree, shown in Fig. \ref{fig:plots 125028-HR}, does not form a triple BH system, because the first BHB (BH$_1$ and BH$_3$, like in the LR version)  forms with an extremely high eccentricity and coalesces faster due to GW emission ($t_{\rm{coal}}$ is predicted via Eq. \ref{eq: tcoal peters}).

\begin{figure}
\centering
\begin{subfigure}[a]{0.5\textwidth}
   \includegraphics[width=\textwidth]
{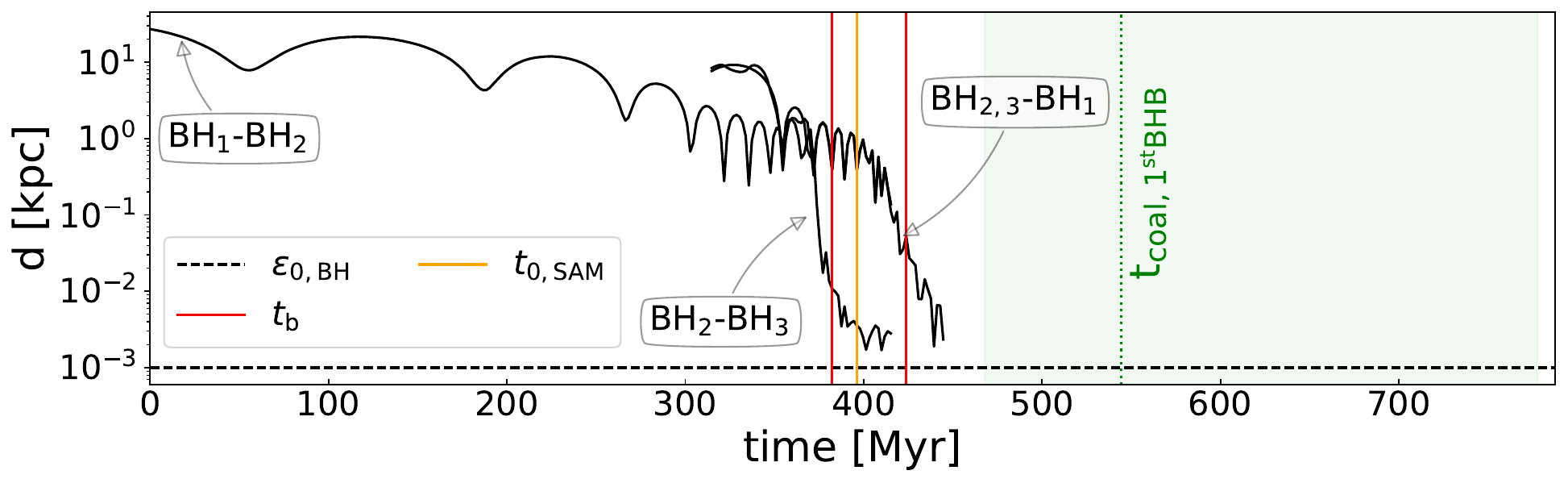}
   \caption{}
   \label{fig:evol 168390-LR}
\end{subfigure}
\hspace{4cm}
   \begin{subfigure}[b]{0.5\textwidth}
   \includegraphics[width=\textwidth]
   {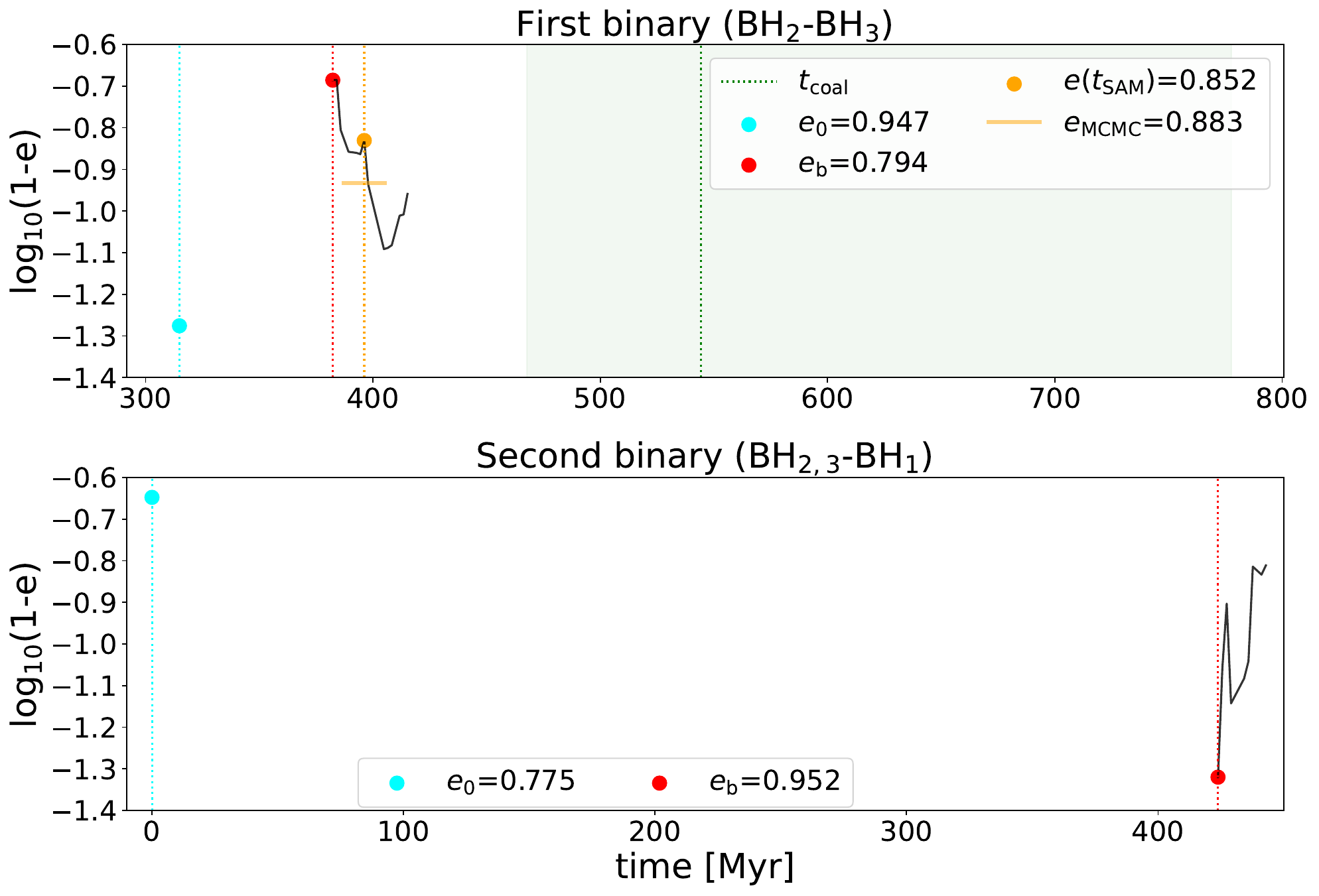}
   \caption{}
   \label{fig:ecc 168390-LR} 
\end{subfigure}
\caption{(a) Merger tree 168390-LR: time evolution of the BHs separation. (b) Evolution of the orbital eccentricity of the first (BH$_{2}$ and BH$_3$) and second (BH$_{2,3}$ and BH$_1$, triplet) binary formed in Tree 168390-LR.}
\label{fig:plots 168390-LR}
\end{figure}

\begin{figure}
\centering
\begin{subfigure}[a]{0.5\textwidth}
   \includegraphics[width=\textwidth]
{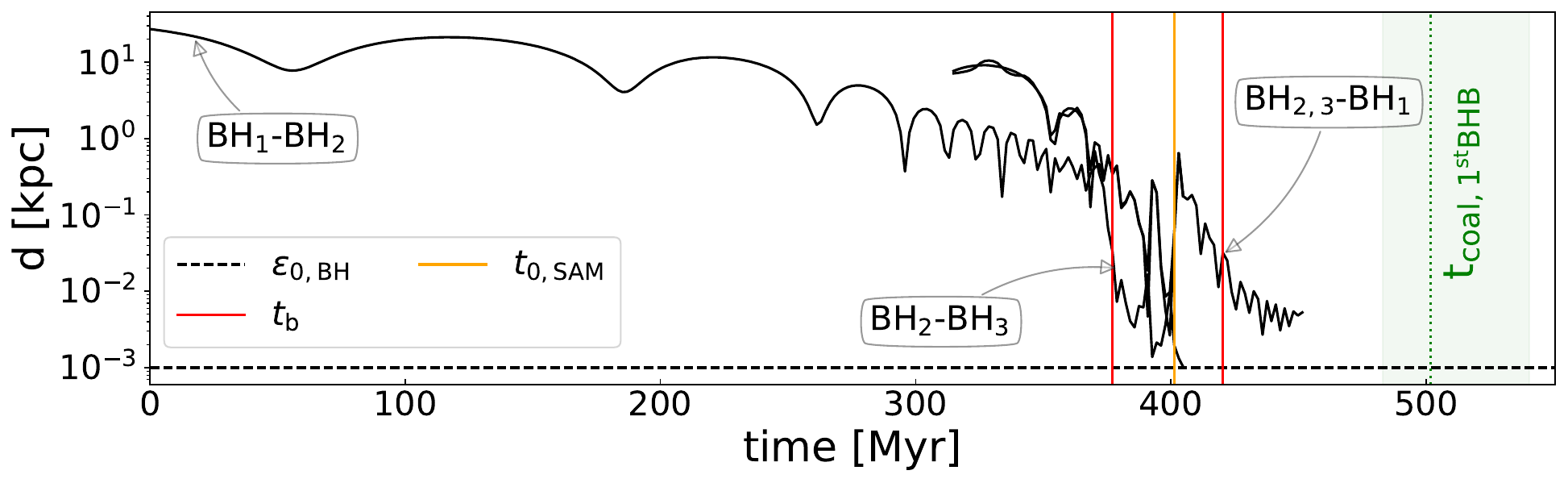}
   \caption{}
   \label{fig:evol 168390-HR}
\end{subfigure}
\hspace{4cm}
   \begin{subfigure}[b]{0.5\textwidth}
   \includegraphics[width=\textwidth]
   {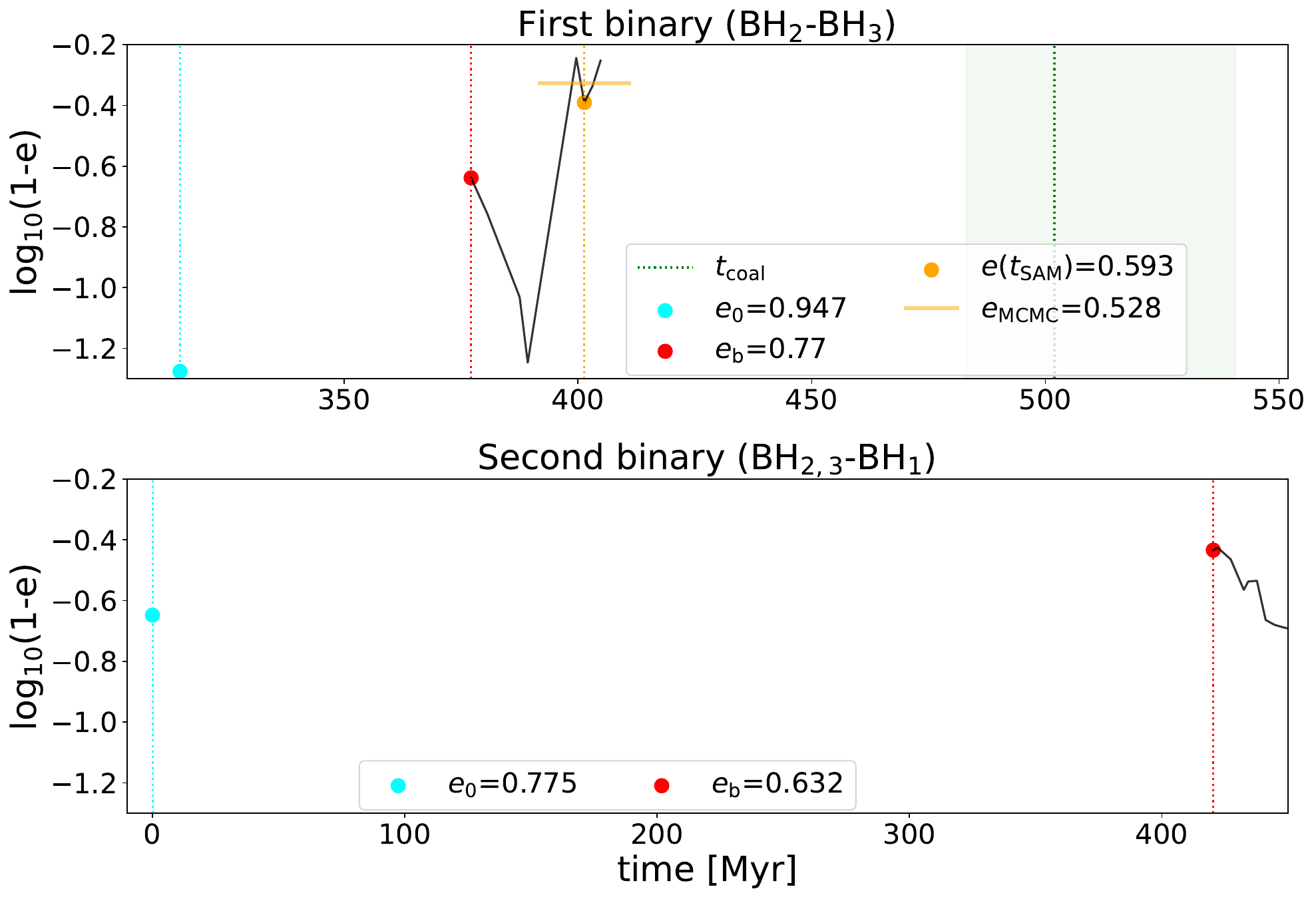}
   \caption{}
   \label{fig:ecc 168390-HR} 
\end{subfigure}
\caption{(a) Merger tree 168390-HR: time evolution of the BHs separation. (b) Evolution of the orbital eccentricity of the first (BH$_{2}$ and BH$_3$) and second (BH$_{2,3}$ and BH$_1$, triplet) binary formed in Tree 168390-HR.}
\label{fig:plots 168390-HR}
\end{figure}

In Fig. \ref{fig:plots 168390-LR} and Fig. \ref{fig:plots 168390-HR} we show data that refer to merger tree 168390, in its LR and HR versions, respectively. This complete 2-merger tree forms a triple BH system in both resolution runs. Like in tree 125028, the third galaxy is initialised at a separation that is smaller than the separation between the primary and secondary galaxy of the first merger, leading BH$_3$ to be involved in the formation of the first binary. Differently from tree 125028, however, BH$_3$ binds with BH$_2$, but the explanation is analogous: BH$_2$ and BH$_3$ are more massive than BH$_1$. In panel (a) of Fig. \ref{fig:plots 168390-HR} we see that in the HR run the three BHs undergo multiple close encounters before the inner BHB merges, while this is not the case in the low-resolution version.

\begin{figure}
\centering
\begin{subfigure}[a]{0.5\textwidth}
   \includegraphics[width=\textwidth]
{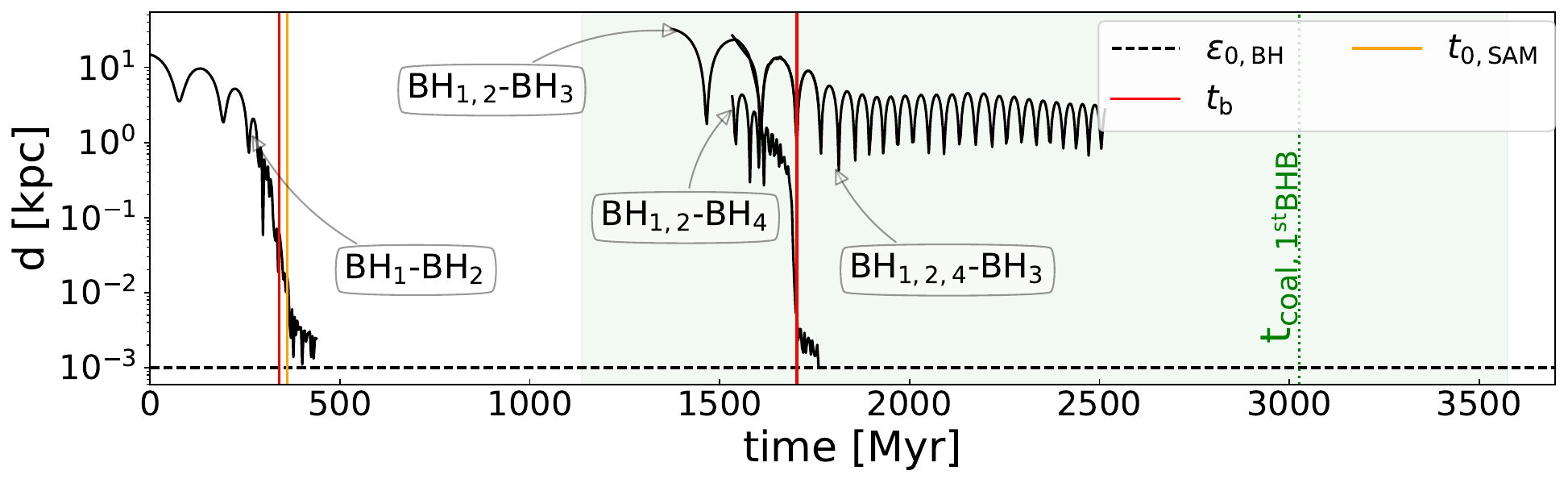}
   \caption{}
   \label{fig:evol 197109-LR}
\end{subfigure}
\hspace{4cm}
\begin{subfigure}[b]{0.5\textwidth}

\includegraphics[width=\textwidth]
   {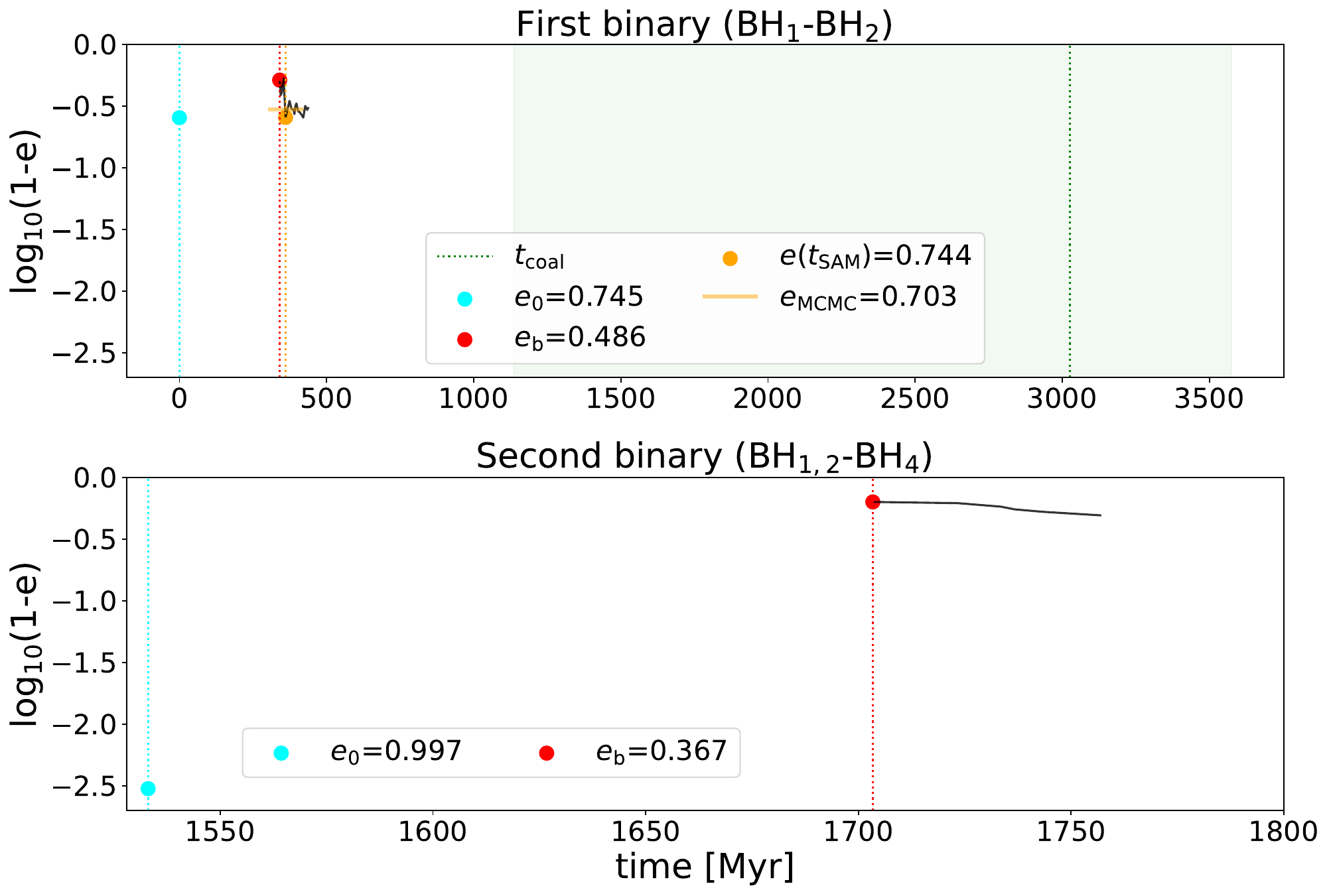}
   \caption{}
   \label{fig:ecc 168390-LR} 
\end{subfigure}
\caption{(a) Merger tree 197109-LR: time evolution of the BHs separation. (b) Evolution of the orbital eccentricity of the first (BH$_{1}$ and BH$_2$), second (BH$_{1,2}$ and BH$_3$, triplet) and third (BH$_{1,2,3}$ and BH$_4$) binary formed in Tree 197109-LR.}
\label{fig:plots 197109-LR}
\end{figure}

Finally, in Fig. \ref{fig:plots 197109-LR} we report data relative to the low-resolution version of tree 197109, the one used as example in the main body of the paper. We note that in this LR run, the first BHB forms with an eccentricity that remains almost constant in the last stages of the {\tt Griffin} evolution. This leads to a small MCMC best-fit value of $K$ (the parameter that refers to the eccentricity growth rate) and thus a long (and uncertain) coalescence time. As a result, the first BHB in this LR version is predicted to coalesce after the formation of the second binary, thus forming a triple BH system.

\begin{figure}
    \centering
    \includegraphics[width=1\linewidth]{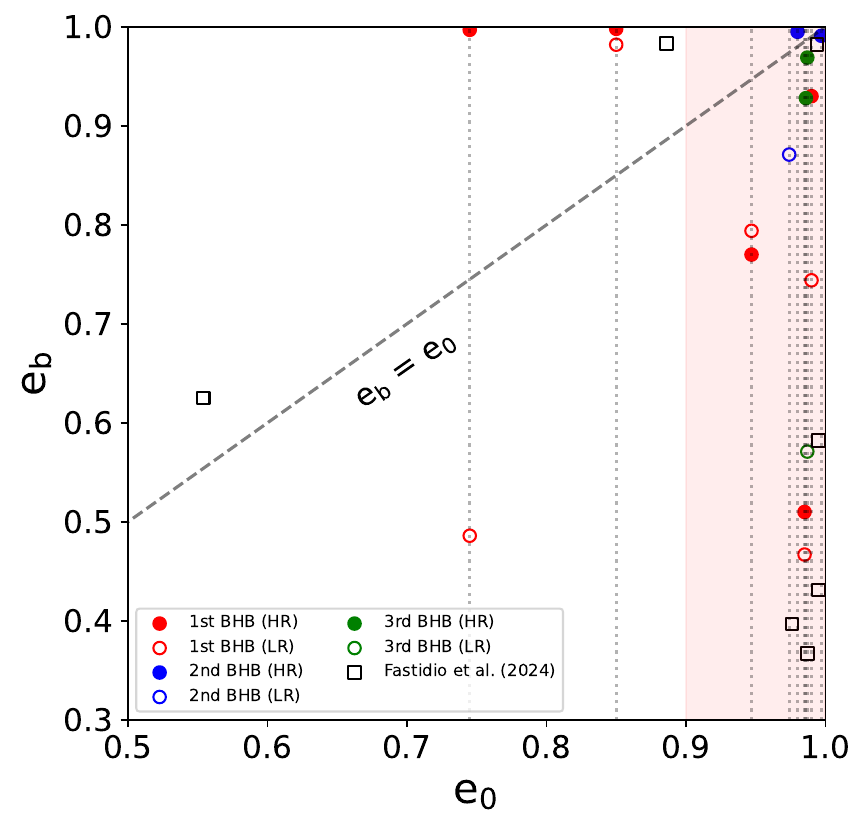}
    \caption{Eccentricity at binary formation ($e_{\rm{b}}$) as a function of the initial galactic merger orbital eccentricity ($e_0$). We use different colours to distinguish between data relative to the first (red), the second (blue) or the third (green) BHB of a tree. Empty and filled dots represent data from the LR and HR runs, respectively. The empty black squares are data from our previous work \citep{2024MNRAS.532..295F}, whose resolution is comparable to our LR runs. The red shaded area highlights the region where $e_0>0.9$}
    \label{fig:correlation no log}
\end{figure}

Fig. \ref{fig:correlation no log} shows the correlation between the initial eccentricity of the galactic mergers and the eccentricity at binary formation, similar to Fig. \ref{fig:ecc corr colors}. This plot is presented on a linear scale, and we have included data from \cite{2024MNRAS.532..295F} (empty black squares) to facilitate comparison.

\newpage

\end{appendix}

\end{document}